\documentclass[10pt,english,5p, sort&compress]{elsarticle}
\usepackage[T1]{fontenc}
\usepackage[latin9]{inputenc}
\usepackage{float}
\usepackage{mathrsfs}
\usepackage{amsmath}
\usepackage{amssymb}
\usepackage{graphicx}

\makeatletter

\providecommand{\tabularnewline}{\\}
\floatstyle{ruled}
\newfloat{algorithm}{tbp}{loa}
\providecommand{\algorithmname}{Algorithm}
\floatname{algorithm}{\protect\algorithmname}

\usepackage{bbold}

\makeatother

\usepackage{babel}
\begin{document}
\begin{frontmatter}

\title{An $\mathcal{O}(N)$ Maxwell solver with improved numerical dispersion
properties}
\begin{abstract}
A Maxwell solver derived from finite element method with $\mathcal{O}(N)$
computing cost is developed to improve the numerical dispersion properties
in relativistic particle-in-cell (PIC) simulations. The correction
of the dispersion relation of the electromagnetic wave is achieved
using the neighboring cells via an iteration scheme without decomposing
into Fourier modes. The local nature of the communication is ideally
suited to massively parallel computer architectures. This Maxwell
solver constrains the Numerical Cherenkov instability (NCI) for the
ultra-relativistic drifting pair plasma in $x$ direction to large
wave vectors for two dimensional grid. The growth rate of NCI is suppressed
by using the low pass filtering.
\end{abstract}
\begin{keyword}
Particle-in-cell, Numerical Cherenkov instability, Maxwell solver
\end{keyword}
\author[1,2]{Yingchao Lu} 
\ead{yclu@lanl.gov}
\address[1]{Theoretical Division, Los Alamos National Laboratory, Los Alamos, New Mexico, 87545, USA}
\address[2]{Department of Physics and Astronomy, Rice University, Houston, Texas 77005, USA}
\author[1]{Chengkun Huang}
\author[1]{Patrick Kilian}
\author[1]{Fan Guo}
\author[1]{Hui Li}
\author[2]{Edison Liang}

\end{frontmatter}

\section{Introduction}

Particle-in-cell (PIC) method\citep{PIC_Birdsall1991} is widely used
for the simulations of plasma dynamics ranging from Laser Plasma Accelerators
(LPAs) to collisionless astrophysical problems. The traditional Yee
scheme\citep{PIC_Yee1966} introduces a resonance between the beam
and electromagnetic (EM) wave, which results in numerical Cherenkov
instability (NCI) \citep{NCI_Godfrey1974}. The resonance is a numerical
artifact, which is due to two fundamental properties in the numerical
schemes: (1) In the Maxwell solver, the dispersion relation of the
vacuum electromagnetic waves deviates from the physical one, i.e.
$\omega=ck$ (we call $\omega=ck$ luminous, $\omega>ck$ superluminous,
$\omega<ck$ subluminous). (2) The dispersion of the drifting plasma
has contribution of aliases\citep{Aliasing_Huang2016}. For example,
in Yee solver the dispersion is superluminous along any axis but luminous
along diagonals and the main beam will intersect at some direction
between these two directions for low $k$. This numerical artifact
has significant impact on the quality and physical interpretation
of relativistic PIC simulations. The numerical instabilities near
$k=0$, once generated, are difficult to be differentiated from the
physics modes, which are also near $k=0$. A numerical scheme without
numerical resonance near $k=0$ or suppressing the NCI growth rate
resolves the numerical artifact. Several approaches have been proposed
to improve the numerical dispersion properties\citep{NCI_Xu2013,Blinne2018,PSATD_Godfrey2014a,PSATD_Godfrey2014b,PSTD_Godfrey2015,NCI_Vay2011,FFTbase_Yu2015,FFTbase_Li2017},
including the optimal choice of time step and spectral solvers.

One approach to minimize the NCI growth rate is achieved by optimal
choice of time step, which was found for both energy conserving and
momentum conserving scheme by using the analytical formula for the
asymptotic growth rate\citep{NCI_Xu2013}. The asymptotic NCI growth
rate is useful for optimizing the numerical dispersion properties
of different interpolation schemes and Maxwell solvers.

Another approach to improve the numerical dispersion is to use the
spatial Fourier transform based methods, such as Pseudo-Spectral Time
Domain (PSTD) algorithms\citep{PSATD_Godfrey2014a,PSATD_Godfrey2014b,PSTD_Godfrey2015}.
Those methods can constrain numerical resonance between EM mode and
beams to large wave vectors, i.e. no resonance near $k=0$. The numerical
instabilities can then be suppressed by applying low-pass filtering\citep{NCI_Vay2011}.
The non-local nature of communication cost and the $\mathcal{O}(N\log N)$
computational cost of the spatial Fourier transform algorithms make
those methods challenging to scale on massively parallel computers.
In the generalized PSTD algorithm\citep{PSTD_Godfrey2015}, the components
of the wavevectors can be replaced by the Fourier transforms of finite
difference approximations to spatial derivatives on a grid, which
reduces the communication cost to local and computational cost to
$\mathcal{O}(N)$. The mixed FD-FFT solver use 1D FFT only\citep{FFTbase_Yu2015}
or high order FDTD in one direction only\citep{FFTbase_Li2017}. However,
the mixed solver nature in different directions requires correction
due to the loss of charge conservation.

In this paper, we aim to develop an alternative numerical scheme on
multi-dimensional grid with superluminous numerical dispersion in
a large range of $k$ including $k=0$, and with $\mathcal{O}(N)$
computing cost and only local communication. The discretized set of
Maxwell equations without lumping, i.e. keeping the averaging operators,
from Eastwood 1991\citep{Eastwood1991} is found to satisfy all our
requirements and has been implemented in this work using the EPOCH
code\citep{EPOCH_Arber2015}. If the charge conserving current deposition
is used, the set of equations also conserve the Gauss's law for electric
field, there is no need for divergence cleaning of electric field
or correction of the current. The divergence-free nature of the magnetic
field is also preserved. In principle, the numerical scheme we develop
here can be implemented in any Yee-grid based PIC code and generalized
to complex geometries. Our numerical scheme modifies Ampere's equations,
instead of modifying the Faraday equation in the finite-difference
time-domain (FDTD) method\citep{Blinne2018}.

This paper is organized as follows. The algorithm for solving Maxwell
equations and the dispersion properties of the numerical electromagnetic
waves in vacuum are described in Sec \ref{sec:Maxwell-Solver}. The
NCI growth rate for drifting pair plasma is discussed in Sec \ref{sec:NCI-growth-rate},
following the analytical technique in Xu 2013\citep{NCI_Xu2013}.
The numerical experiments are carried out to confirm the two dimensional
EM dispersion relation and the NCI growth rate in Sec \ref{sec:Numerical-experiments}.

\section{Maxwell Solver\label{sec:Maxwell-Solver}}

In Eastwood \citep{Eastwood1991}, the coupled relativistic Vlasov-Maxwell
set of equations are derived using finite elements in both space and
time. Their finite element derivations are suitable for complex geometries.
In this paper, we solve the Vlasov-Maxwell equation on uniform rectangular
grid by using Faraday and Ampere's equations from Eastwood, without
modifying other modules from the usual finite difference PIC, i.e.,
force interpolation, current deposition and particle pusher. The discretized
Faraday equations from Eastwood 1991 can be generalized to 3D
\begin{align}
\frac{1}{c}D_{t}B_{x} & =-\big[D_{y}E_{z}-D_{z}E_{y}\big]\label{eq:faraday_x}\\
\frac{1}{c}D_{t}B_{y} & =-\big[D_{z}E_{x}-D_{x}E_{z}\big]\label{eq:faraday_y}\\
\frac{1}{c}D_{t}B_{z} & =-\big[D_{x}E_{y}-D_{y}E_{x}\big]\label{eq:faraday_z}
\end{align}
and the Ampere's equations are 
\begin{align}
\frac{1}{c}D_{t}(\alpha_{yz}E_{x}) & =\big[D_{y}(\alpha_{tz}B_{z})-D_{z}(\alpha_{ty}B_{y})\big]-\frac{4\pi}{c}j_{x}\label{eq:ampere_x}\\
\frac{1}{c}D_{t}(\alpha_{zx}E_{y}) & =\big[D_{z}(\alpha_{tx}B_{x})-D_{x}(\alpha_{tz}B_{z})\big]-\frac{4\pi}{c}j_{y}\label{eq:ampere_y}\\
\frac{1}{c}D_{t}(\alpha_{xy}E_{z}) & =\big[D_{x}(\alpha_{ty}B_{y})-D_{y}(\alpha_{tx}B_{x})\big]-\frac{4\pi}{c}j_{z}\label{eq:ampere_z}
\end{align}
and the Gauss's law for electric field and magnetic field 
\begin{align}
D_{x}(\alpha_{yz}E_{x})+D_{y}(\alpha_{zx}E_{y})+D_{z}(\alpha_{xy}E_{z}) & =4\pi\rho\label{eq:gauss_E}\\
D_{x}B_{x}+D_{y}B_{y}+D_{z}B_{z} & =0\label{eq:gauss_B}
\end{align}
where difference operators $D_{x}$, $D_{y}$, $D_{z}$, $D_{t}$
and averaging operators $\alpha_{x}$, $\alpha_{y}$, $\alpha_{z}$,
$\alpha_{t}$ respectively act on the spatial indices $k$, $l$,
$m$ and time step index $n$:
\begin{align}
D_{x}f_{k,l,m,n} & =\frac{f_{k+1/2,l,m,n}-f_{k-1/2,l,m,n}}{\Delta x}\\
\alpha_{x}f_{k,l,m,n} & =\frac{f_{k+1,l,m,n}+4f_{k,l,m,n}+f_{k-1,l,m,n}}{6}
\end{align}
and similarly for $D_{y}$ and $\alpha_{y}$ on index $l$, and for
$D_{z}$ and $\alpha_{z}$ on index $m$, and for $D_{t}$ and $\alpha_{t}$
on the index $n$. The product of two averaging operators $\alpha_{y}$
and $\alpha_{z}$ is abbreviated as $\alpha_{yz}$, similarly for
averaging operators in other directions. All the indices can be integers
or half integers, depending on the staggered directions of the grid
and time step. As derived from finite element method\citep{Eastwood1991},
the averaging stencil is $(\frac{1}{6},\frac{2}{3},\frac{1}{6})$
for spatial grid in all three spatial dimensions and also for time
step. The variables for electric field $\overrightarrow{E}$, magnetic
field $\overrightarrow{B}$ and current density \textbf{$\overrightarrow{\jmath}$}
are defined either at grid point or has a half-grid offset, on integer
or half-integer time step, as same as in the Yee scheme. The whole
set of discrete variables is 
\begin{align*}
 & E_{x}^{k+1/2,l,m,n},E_{y}^{k,l+1/2,m,n},E_{z}^{k,l,m+1/2,n},\\
 & B_{x}^{k,l+1/2,m+1/2,n+1/2},B_{y}^{k+1/2,l,m+1/2,n+1/2},\\
 & B_{z}^{k+1/2,l+1/2,m,n+1/2},\\
 & j_{x}^{k+1/2,l,m,n},j_{y}^{k,l+1/2,m,n},j_{z}^{k,l,m+1/2,n}\\
 & \rho^{k,l,m,n+1/2}
\end{align*}
where $k$, $l$, $m$, $n$ are integers. By lumping the averaging
operators, $\alpha_{x},\alpha_{y},\alpha_{z},\alpha_{t}\to1$ in Eqs(\ref{eq:faraday_x})
to Eq(\ref{eq:ampere_z}), we recover the discretized equations for
Yee solver. In the numerical scheme we use in this work, we keep those
averaging operators instead of lumping them in order to achieve the
desired numerical dispersion properties.

By performing Fourier transform on Eqs(\ref{eq:faraday_x}) to (\ref{eq:ampere_z}),
assuming that all the variables have $e^{i(k_{x}x+k_{y}y+k_{z}z-\omega t)}$
form, the difference and averaging operators can be written in terms
of frequency $\omega$ or wave vector $\overrightarrow{k}$
\begin{align}
\alpha_{x} & =\phantom{-i}\frac{2+\cos k_{x}\Delta x}{3}=1-\frac{2}{3}\sin^{2}\frac{k_{x}\Delta x}{2}\label{eq:alphax}\\
D_{x} & =\phantom{-}i\frac{\sin(k_{x}\Delta x/2)}{\Delta x/2}\\
D_{t} & =-i\frac{\sin(\omega\Delta t/2)}{\Delta t/2}
\end{align}
and similarly for $\alpha_{t},\alpha_{y},\alpha_{z},D_{y},D_{z}$,
where $\Delta x$ is the grid size in $x$ direction, $\Delta t$
is the time step. The Eqs(\ref{eq:faraday_x}) to (\ref{eq:ampere_z})
in Fourier space can be written as 
\begin{align}
[\overrightarrow{k}]\times\overrightarrow{E} & =\frac{[\omega]}{c}\overrightarrow{B}\label{eq:faraday_fourier}\\{}
[\overrightarrow{k}]\times(\mathscr{A}_{B}\overrightarrow{B} & )=-\frac{[\omega]}{c}\mathscr{A}_{E}\overrightarrow{E}-\frac{4\pi i}{c}\overrightarrow{\jmath}\label{eq:ampere_fourier}
\end{align}
where 
\begin{align}
[k]_{i} & =\frac{\sin(k_{i}\Delta x_{i}/2)}{\Delta x_{i}/2},\quad i=1,2,3\label{eq:braket_k}\\{}
[\omega] & =\frac{\sin(\omega\Delta t/2)}{\Delta t/2},
\end{align}
\begin{align}
\mathscr{A}_{B} & =\begin{pmatrix}\alpha_{tx} & 0 & 0\\
0 & \alpha_{ty} & 0\\
0 & 0 & \alpha_{tz}
\end{pmatrix},\quad\mathrm{and}\\
\mathscr{A}_{E} & =\begin{pmatrix}\alpha_{yz} & 0 & 0\\
0 & \alpha_{zx} & 0\\
0 & 0 & \alpha_{xy}
\end{pmatrix}
\end{align}
By multiplying $[\omega]$ on Eq(\ref{eq:ampere_fourier}) and using
Eq(\ref{eq:faraday_fourier}) to eliminate $\overrightarrow{B}$,
we get
\begin{equation}
[\omega]^{2}\mathscr{A}_{E}\overrightarrow{E}+c^{2}[\overrightarrow{k}]\times\mathscr{A}_{B}([\overrightarrow{k}]\times\overrightarrow{E})=-4\pi i[\omega]\overrightarrow{\jmath}\label{eq:EJ}
\end{equation}
We define the vacuum part of dielectric tensor by $\overleftrightarrow{\epsilon}^{(\mathrm{vac})}\cdot\overrightarrow{E}=[\omega]^{2}\mathscr{A}_{E}\overrightarrow{E}+c^{2}[\overrightarrow{k}]\times\mathscr{A}_{B}([\overrightarrow{k}]\times\overrightarrow{E})$,
then the matrix elements of $\overleftrightarrow{\epsilon}^{(\mathrm{vac})}$
can be written down
\begin{align}
\epsilon_{xx}^{(\mathrm{vac})} & =\alpha_{yz}[\omega]^{2}-\alpha_{ty}c^{2}[k]_{z}^{2}-\alpha_{tz}c^{2}[k]_{y}^{2}\label{eq:eps_em_xx}\\
\epsilon_{yy}^{(\mathrm{vac})} & =\alpha_{zx}[\omega]^{2}-\alpha_{tz}c^{2}[k]_{x}^{2}-\alpha_{tx}c^{2}[k]_{z}^{2}\\
\epsilon_{zz}^{(\mathrm{vac})} & =\alpha_{xy}[\omega]^{2}-\alpha_{tx}c^{2}[k]_{y}^{2}-\alpha_{ty}c^{2}[k]_{x}^{2}\\
\epsilon_{xy}^{(\mathrm{vac})} & =\epsilon_{yx}^{(\mathrm{vac})}=\alpha_{tz}c^{2}[k]_{x}[k]_{y}\\
\epsilon_{yz}^{(\mathrm{vac})} & =\epsilon_{zy}^{(\mathrm{vac})}=\alpha_{tx}c^{2}[k]_{y}[k]_{z}\\
\epsilon_{zx}^{(\mathrm{vac})} & =\epsilon_{xz}^{(\mathrm{vac})}=\alpha_{ty}c^{2}[k]_{z}[k]_{x}\label{eq:eps_em_zx}
\end{align}
By lumping the averaging operators, i.e. letting $\mathscr{A}_{E}\to\mathbb{1}$
and $\mathscr{A}_{B}\to\mathbb{1}$ in Eq(\ref{eq:EJ}), the vacuum
dielectric tensor for Yee solver is recovered.

It is derived in Appendix \ref{sec:Conservation} that Eq(\ref{eq:gauss_E})
is conserved automatically if it is fulfilled at the initial condition
and a charge conserving current deposition scheme is used. And Eq(\ref{eq:gauss_B})
is conserved if it is fulfilled at the initial condition.

\subsection{Dispersion relation for EM waves in vacuum}

In vacuum we have zero current $\overrightarrow{j}=0$, the dispersion
relation of EM waves can be obtained by calculating the determinant
of the dielectric tensor $\overleftrightarrow{\epsilon}^{(\mathrm{vac})}$
\begin{align}
\det\overleftrightarrow{\epsilon}^{(\mathrm{vac})} & =(\alpha_{xyzt})^{2}[\omega]^{2}\big(\alpha_{t}^{-1}\frac{[\omega]^{2}}{c^{2}}\nonumber \\
 & \qquad-\alpha_{x}^{-1}[k]_{x}^{2}+\alpha_{y}^{-1}[k]_{y}^{2}+\alpha_{z}^{-1}[k]_{z}^{2}\big)^{2}
\end{align}
The $\omega=0$ mode represents the background field while the other
root of $\omega$ represents the EM mode. Thus we obtain the vacuum
EM wave dispersion relation
\begin{equation}
\alpha_{t}^{-1}\frac{[\omega]^{2}}{c^{2}}=\alpha_{x}^{-1}[k]_{x}^{2}+\alpha_{y}^{-1}[k]_{y}^{2}+\alpha_{z}^{-1}[k]_{z}^{2}\label{eq:dispersion-em-original}
\end{equation}
We define the function $G(\theta)$ to be
\begin{equation}
G(\theta)=\frac{\sin^{2}\theta}{1-\frac{2}{3}\sin^{2}\theta}\label{eq:G}
\end{equation}
Then Eq(\ref{eq:dispersion-em-original}) becomes
\begin{equation}
\frac{G(\frac{\omega\Delta t}{2})}{c^{2}\Delta t^{2}}=\frac{G(\frac{k_{x}\Delta x}{2})}{\Delta x^{2}}+\frac{G(\frac{k_{y}\Delta y}{2})}{\Delta y^{2}}+\frac{G(\frac{k_{z}\Delta z}{2})}{\Delta z^{2}}\label{eq:dispersion-em-F-3d}
\end{equation}
For numerical stability, the frequency $\omega$ must be real which,
by Eq(\ref{eq:dispersion-em-F-3d}) implies the same time step constraint
as Yee solver
\begin{equation}
c\Delta t_{\mathrm{3D}}\le c\Delta t_{\mathrm{3D,CFL}}=(1/\Delta x^{2}+1/\Delta y^{2}+1/\Delta z^{2})^{-1/2}\label{eq:dt_cfl_3d}
\end{equation}
The detail derivation of Eq(\ref{eq:dt_cfl_3d}) is in Appendix \ref{sec:Derivation-inequties}.
For $k\to0$, we expand the phase speed $\omega/k$ to second order
in $k$
\begin{align}
\omega/k & =c\bigg[1+\frac{1}{24}(\frac{k_{x}^{4}\Delta x^{2}+k_{y}^{4}\Delta y^{2}+k_{z}^{4}\Delta z^{2}}{k^{2}}-k^{2}c^{2}\Delta t^{2})\nonumber \\
 & \phantom{=c\bigg[1}+\mathcal{O}(k^{4})\bigg]\label{eq:phase-speed-3d}
\end{align}

From Eq(\ref{eq:phase-speed-3d}) we derive in Appendix \ref{sec:Derivation-inequties}
that the EM mode for sufficiently small $k$ always has phase speed
larger than or equal to $c$, i.e. the speed of light, so the resonance
between EM mode and unaliased plasma beam with any physical speed
$v<c$ cannot be located near $k=0$.

\begin{figure*}
\includegraphics[scale=0.4]{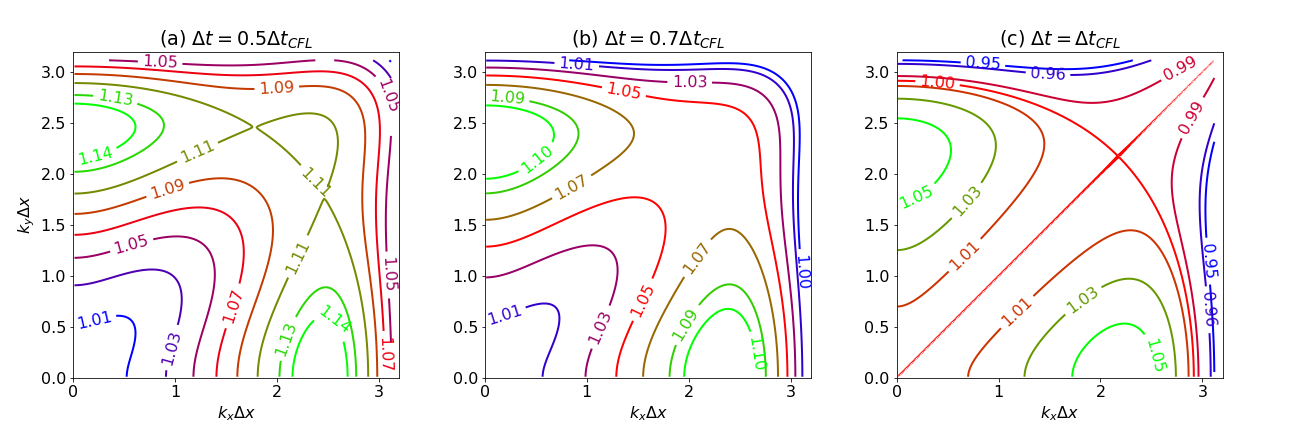}

\includegraphics[scale=0.4]{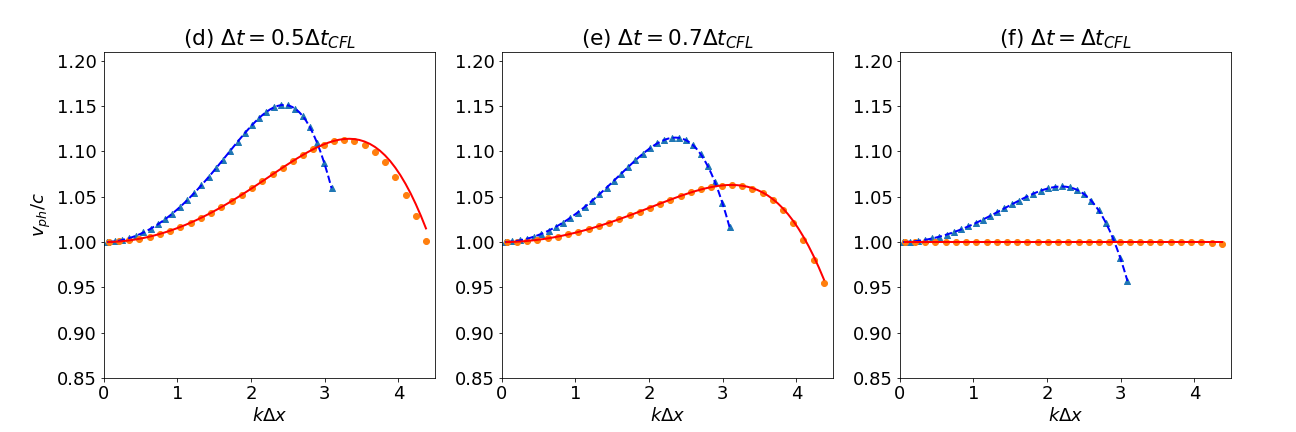}

\caption{Numerical properties for the phase velocities of our solver. Subfigures
(a) to (c) are two dimensional contours of phase speed normalized
by speed of light in the numerical dispersion given by Eq(\ref{eq:dispersion-em-F-2d}),
for the $\Delta x=\Delta y$ case, and for different time steps $\Delta t$.
We only plot the region where $k_{x}>0$ and $k_{y}>0$ because the
dispersion relation is symmetric under the transform $k_{x}\to-k_{x}$
or $k_{y}\to-k_{y}$. For $\Delta t=\Delta t_{\mathrm{CFL}}$ case,
the phase velocity is always $c$ along the diagonal. Subfigures (d)
to (f) are the numerical phase velocities along the $k_{x}$ axis
and along the diagonal. The numerical dispersion relation are tested
by taking the Fourier transform of $E_{z}$ in the simulations against
the dispersion relation in Eq(\ref{eq:dispersion-em-F-2d}). \label{fig:dispersion-EM2d}}
\end{figure*}

For a two-dimensional grid, we can write down the dispersion relation
of the EM mode, the CFL limit of time step and the expansion of phase
speed in the similar way
\begin{align}
\omega/k & =c\bigg[1+\frac{1}{24}(\frac{k_{x}^{4}\Delta x^{2}+k_{y}^{4}\Delta y^{2}}{k^{2}}-k^{2}c^{2}\Delta t^{2})\nonumber \\
 & \phantom{=c\bigg[1}+\mathcal{O}(k^{4})\bigg]\label{eq:phase-speed-2d}
\end{align}
\begin{equation}
\frac{G(\frac{\omega\Delta t}{2})}{c^{2}\Delta t^{2}}=\frac{G(\frac{k_{x}\Delta x}{2})}{\Delta x^{2}}+\frac{G(\frac{k_{y}\Delta y}{2})}{\Delta y^{2}}\label{eq:dispersion-em-F-2d}
\end{equation}
\begin{equation}
c\Delta t_{\mathrm{2D}}\le c\Delta t_{\mathrm{2D,CFL}}=(1/\Delta x^{2}+1/\Delta y^{2})^{-1/2}
\end{equation}
For the the $\Delta x=\Delta y$ case, we plot the phase velocity
in $k$ space for different values of time step in Fig. \ref{fig:dispersion-EM2d}.

In the 2D or 3D case for $\Delta t=\Delta t_{\mathrm{CFL}}$, the
phase speed is constant and equals to $c$ along the diagonal of the
first Brillouin zone, i.e. $|k_{x}|\Delta x=|k_{y}|\Delta y=|k_{z}|\Delta z$
in 3D case and $|k_{x}|\Delta x=|k_{y}|\Delta y$ in 2D case. For
$\Delta t\ne\Delta t_{\mathrm{CFL}}$ or off-diagonal direction, the
second order term in Eq(\ref{eq:phase-speed-3d}) and (\ref{eq:phase-speed-2d})
is always larger than zero, thus $\omega/k$ is larger than $c$ for
sufficiently small $k$.

\subsection{Numerical iteration scheme}

\subsubsection{Equations for three dimensional case}

Updating $\overrightarrow{B}$ is fully explicit by using Eqs(\ref{eq:faraday_x})
to (\ref{eq:faraday_z}). However, Eqs(\ref{eq:ampere_x}) to (\ref{eq:ampere_z})
contain the time averaging operator on the L.H.S., which require the
field at a previous step and a future step. This requirement can be
eliminated by substituting Eq(\ref{eq:faraday_x}) to (\ref{eq:faraday_z})
into the averaging operators. For example, the time averaging of $B_{z}$
can be written as 
\begin{align}
\alpha_{t}B_{z}^{n+1/2} & =\frac{B_{z}^{n-1/2}+4B_{z}^{n+1/2}+B_{z}^{n+3/2}}{6}\nonumber \\
 & =\frac{B_{z}^{n+1/2}-\Delta t(D_{t}B_{z})^{n}}{6}+\frac{2}{3}B_{z}^{n+1/2}\nonumber \\
 & \qquad+\frac{B_{z}^{n+1/2}+\Delta t(D_{t}B_{z})^{n+1}}{6}\nonumber \\
 & =B_{z}^{n+1/2}+\frac{\Delta t}{6}(D_{t}B_{z})^{n+1}-\frac{\Delta t}{6}(D_{t}B_{z})^{n}\nonumber \\
 & =B_{z}^{n+1/2}+\frac{c\Delta t}{6}(-D_{x}E_{y}+D_{y}E_{x})^{n+1}\nonumber \\
 & \qquad-\frac{c\Delta t}{6}(-D_{x}E_{y}+D_{y}E_{x})^{n}\label{eq:alphat_bz}
\end{align}
where we used Eq(\ref{eq:faraday_z}) in the derivation. Similarly
we can write down the time averaging of $B_{y}$ as 
\begin{align}
\alpha_{t}B_{y}^{n+1/2} & =B_{y}^{n+1/2}+\frac{c\Delta t}{6}(-D_{z}E_{x}+D_{x}E_{z})^{n+1}\nonumber \\
 & \qquad-\frac{c\Delta t}{6}(-D_{z}E_{x}+D_{x}E_{z})^{n}\label{eq:alphat_by}
\end{align}
Substituting Eq(\ref{eq:alphat_bz}) and (\ref{eq:alphat_by}) into
the time averaging terms in Eq(\ref{eq:ampere_x}), we obtain
\begin{align}
 & \alpha_{yz}\frac{E_{x}^{n+1}-E_{x}^{n}}{c\Delta t}\nonumber \\
= & D_{y}\alpha_{z}[B_{z}^{n+1/2}+\frac{c\Delta t}{6}(-D_{x}E_{y}+D_{y}E_{x})^{n+1}\nonumber \\
 & \qquad-\frac{c\Delta t}{6}(-D_{x}E_{y}+D_{y}E_{x})^{n}]\nonumber \\
 & -D_{z}\alpha_{y}[B_{y}^{n+1/2}+\frac{c\Delta t}{6}(-D_{z}E_{x}+D_{x}E_{z})^{n+1}\nonumber \\
 & \qquad-\frac{c\Delta t}{6}(-D_{z}E_{x}+D_{x}E_{z})^{n}]\nonumber \\
 & -\frac{4\pi}{c}j_{x}^{n+1/2}\label{eq:derivation-scheme}
\end{align}
where $n$ is the index for $n$th time step. If we define the auxiliary
variable $E_{x}^{\prime}$ such that 
\begin{align}
D_{t}E_{x}^{\prime} & =D_{t}[\alpha_{yz}E_{x}\nonumber \\
 & \qquad-\frac{c^{2}\Delta t^{2}}{6}D_{y}\alpha_{z}(-D_{x}E_{y}+D_{y}E_{x})\nonumber \\
 & \qquad+\frac{c^{2}\Delta t^{2}}{6}D_{z}\alpha_{y}(-D_{z}E_{x}+D_{x}E_{z})]\nonumber \\
 & =D_{t}[(\alpha_{yz}-\frac{c^{2}\Delta t^{2}}{6}D_{y}^{2}\alpha_{z}-\frac{c^{2}\Delta t^{2}}{6}D_{z}^{2}\alpha_{y})E_{x}\nonumber \\
 & \qquad+\frac{c^{2}\Delta t^{2}}{6}D_{x}D_{y}\alpha_{z}E_{y}+\frac{c^{2}\Delta t^{2}}{6}D_{x}D_{z}\alpha_{y}E_{z}]
\end{align}
then we can simplify Eq(\ref{eq:derivation-scheme}) to be
\begin{equation}
\frac{1}{c}D_{t}E_{x}^{\prime}=D_{y}(\alpha_{z}B_{z})-D_{z}(\alpha_{y}B_{y})-\frac{4\pi}{c}j_{x}\label{eq:ampere_aux_x}
\end{equation}
The equations for the auxiliary variables are similar in $y$ and
$z$ directions
\begin{align}
\frac{1}{c}D_{t}E_{y}^{\prime} & =D_{z}(\alpha_{x}B_{x})-D_{x}(\alpha_{z}B_{z})-\frac{4\pi}{c}j_{y}\label{eq:ampere_aux_y}\\
\frac{1}{c}D_{t}E_{z}^{\prime} & =D_{x}(\alpha_{y}B_{y})-D_{y}(\alpha_{x}B_{x})-\frac{4\pi}{c}j_{z}\label{eq:ampere_aux_z}
\end{align}
And the change of electric field vector from $n$th step to $(n+1)$th
step $\Delta\overrightarrow{E}=(\Delta E_{x},\Delta E_{y},\Delta E_{z})$
and the change of the auxiliary vector $\Delta\overrightarrow{E}^{\prime}=(\Delta E_{x}^{\prime},\Delta E_{y}^{\prime},\Delta E_{z}^{\prime})$
satisfy the linear transform

\begin{equation}
(\mathbb{1}-\overleftrightarrow{A})\cdot\Delta\overrightarrow{E}=\Delta\overrightarrow{E}^{\prime}\label{eq:transform-e-e_aux}
\end{equation}
 The matrix elements of $A$ can then be written down

\begin{align}
A_{xx} & =\frac{\frac{c^{2}\Delta t^{2}}{\Delta y^{2}}-1}{6}\bar{D}_{yy}+\frac{\frac{c^{2}\Delta t^{2}}{\Delta z^{2}}-1}{6}\bar{D}_{zz}\nonumber \\
 & \qquad+\frac{\frac{c^{2}\Delta t^{2}}{\Delta y^{2}}+\frac{c^{2}\Delta t^{2}}{\Delta z^{2}}-1}{36}\bar{D}{}_{yyzz}\label{eq:matrixA_1st}\\
A_{yy} & =\frac{\frac{c^{2}\Delta t^{2}}{\Delta z^{2}}-1}{6}\bar{D}_{zz}+\frac{\frac{c^{2}\Delta t^{2}}{\Delta x^{2}}-1}{6}\bar{D}_{xx}\nonumber \\
 & \qquad+\frac{\frac{c^{2}\Delta t^{2}}{\Delta z^{2}}+\frac{c^{2}\Delta t^{2}}{\Delta x^{2}}-1}{36}\bar{D}{}_{zzxx}\\
A_{zz} & =\frac{\frac{c^{2}\Delta t^{2}}{\Delta x^{2}}-1}{6}\bar{D}_{xx}+\frac{\frac{c^{2}\Delta t^{2}}{\Delta y^{2}}-1}{6}\bar{D}_{yy}\nonumber \\
 & \qquad+\frac{\frac{c^{2}\Delta t^{2}}{\Delta x^{2}}+\frac{c^{2}\Delta t^{2}}{\Delta y^{2}}-1}{36}\bar{D}{}_{xxyy}
\end{align}

\begin{align}
A_{yz} & =A_{zy}=-\frac{c^{2}\Delta t^{2}}{6\Delta y\Delta z}\bar{D}_{yz}(1+\frac{\bar{D}_{xx}}{6})\\
A_{zx} & =A_{xz}=-\frac{c^{2}\Delta t^{2}}{6\Delta z\Delta x}\bar{D}_{zx}(1+\frac{\bar{D}_{yy}}{6})\\
A_{xy} & =A_{yx}=-\frac{c^{2}\Delta t^{2}}{6\Delta x\Delta y}\bar{D}_{xy}(1+\frac{\bar{D}_{zz}}{6})\label{eq:matrixA_last}
\end{align}
where we define the dimensionless operator $\bar{D}_{x}$ to be the
value difference between the right cell and left cell in $x$ direction
\begin{equation}
\bar{D}_{x}f=\Delta xD_{x}f=f_{i+1/2}-f_{i-1/2}
\end{equation}
and similarly for $y$ and $z$ directions. $\bar{D}$ with multiple
indexes represents the abbreviation of multiplication, e.g. $\bar{D}_{xx}=\bar{D}_{x}\bar{D}_{x}$
and $\bar{D}_{xxyy}=\bar{D}_{x}\bar{D}_{x}\bar{D}_{y}\bar{D}_{y}$.

To update the electric field, we first compute the change of auxiliary
field explicitly by using Eqs(\ref{eq:ampere_aux_x}) to (\ref{eq:ampere_aux_z}).
Then the change of the electric field can be obtained by solving Eq(\ref{eq:transform-e-e_aux})
and the electric field is updated simply by the increment $\overrightarrow{E}^{n+1}=\overrightarrow{E}^{n}+\Delta\overrightarrow{E}$.
To solve for $\Delta\overrightarrow{E}$ by Eq(\ref{eq:transform-e-e_aux}),
we propose to use the iteration method listed in Alg \ref{alg:iteration}.
We prove in Appendix \ref{sec:Derivation-of-norm} that the infinity
norm of matrix $\overleftrightarrow{A}$ is always less unity in 1D
and 2D case. For 3D case, the infinity norm of $A$ is less than unity
if $\Delta t\le\frac{\min(\Delta x,\Delta y,\Delta z)}{\sqrt{2}c}$.
In the 3D and $\Delta x=\Delta y=\Delta z$ case the infinity norm
of matrix $A$ is always less than $1$ because $\Delta t\le\frac{\Delta x}{\sqrt{3}c}$.
Thus the iteration in Alg \ref{alg:iteration} always converges to
the exact solution. Further more the infinity norm or eigenvalues
of matrix $\overleftrightarrow{A}$ does not depend on the number
of cells on the mesh, which implies that the speed of convergence
is not dependent on the size of the problem.

\begin{figure*}
\includegraphics[scale=0.5]{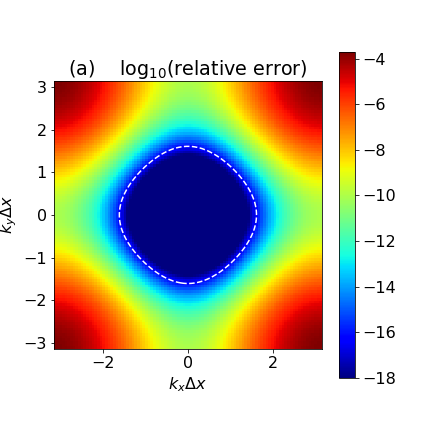}\includegraphics[scale=0.5]{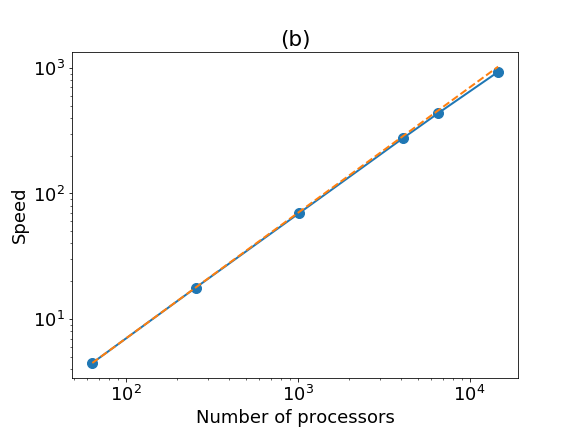}

\caption{The visualization for the accuracy and scaling performance of iterations
in Alg (\ref{alg:iteration}). (a) The logarithm of relative error
for numerical dispersion relation given by Eq(\ref{eq:dispersion_error})
after $m=20$ iterations, for $\Delta x=\Delta y$ and $\Delta t=\Delta t_{\mathrm{CFL}}$.
Zero error means the numerical dispersion relation after iterations
is given by Eq(\ref{eq:dispersion-em-F-2d}). The largest error is
$2\times10^{-4}$ for the edge of the Brillouin zone, i.e. $|k_{x}\Delta x|=|k_{y}\Delta y|=\pi$.
The error is smaller than $10^{-16}$ (the precision for double-precision
floating-point) for the modes near $k=0$ inside area closed by the
white dashed curve. (b) The results for weak scaling of the field
solver, plotted is the curve for speed($=\frac{\mathrm{number\ of\ processors}}{\mathrm{runtime\ in\ seconds}}$)
for 100 time steps vs. the number of processors $N_{\mathrm{proc}}$.
Each processor resolves a $256\times256$ grid. \label{fig:performance}}
\end{figure*}

For several reasons, we propose to use a predetermined number for
the number of iterations instead of determine $|\Delta\overrightarrow{E}_{(m+1)}-\Delta\overrightarrow{E}_{(m)}|<\varepsilon$
during the iterations, (1) calculating the norm $|\Delta\overrightarrow{E}_{(m+1)}-\Delta\overrightarrow{E}_{(m)}|$
and determine $|\Delta\overrightarrow{E}_{(m+1)}-\Delta\overrightarrow{E}_{(m)}|<\varepsilon$
requires a global reduction and broadcasting among processors (2)
keeping fixed number of iterations can keep the consistency of the
error of dispersion relation throughout all time steps (3) fixed number
of iterations is sufficient for reducing the error, e.g. Fig. \ref{fig:performance}(a)
shows that for $\Delta x=\Delta y$ and $\Delta t=\Delta t_{\mathrm{CFL}}$
case, 20 iterations are sufficient to keep the low error for numerical
dispersion relation, especially for the modes near $k=0$. The precondition
for the matrix $\overleftrightarrow{A}$ or relaxation method can
be used to accelerate the convergence and will be subjects of future
reports.

The computational cost of the solver is $\mathcal{O}(N)$ where $N$
is the number of cells. For each iteration, the new quantity $\Delta\overrightarrow{E}_{(m+1)}$
(subscript in parentheses for iteration step, not time step or spatial
grid) only depends on the old quantity $\Delta\overrightarrow{E}_{(m)}$
at the nearest neighbor and next nearest neighbor cells, so that in
the domain decomposed mesh, the communication is local, i.e. communication
only needed to get the quantities at the adjacent cells. Our Maxwell
solver needs slightly more memory than Yee solver due to the auxiliary
field variables. Comparing to the global FFT-based solvers, where
the computing cost is $\mathcal{O}(N\log N)$ and the requirement
for communication is non-local, our algorithm is less computational
expensive and more scalable. In Fig. \ref{fig:performance}(b), we
show the weak-scaling performance of our solver by testing without
particles, for 100 time steps. Each processor resolves a $256\times256$
grid, regardless of what number of processors $N_{\mathrm{proc}}$
is used. The runtime only increases from 14.32 seconds for $N_{\mathrm{proc}}=64$
to 15.75 seconds for $N_{\mathrm{proc}}=14641$, which implies that
the slowdown of the speed is only $9\%$ and the scaling of the solver
is close to $\mathcal{O}(N)$.

\begin{algorithm}
$\Delta\overrightarrow{E}_{(0)}=\Delta\overrightarrow{E}^{\prime}$

$m=0$

while $m<m_{\mathrm{max}}$

\qquad{}$\Delta\overrightarrow{E}_{(m+1)}=\Delta\overrightarrow{E}^{\prime}+\overleftrightarrow{A}\cdot\Delta\overrightarrow{E}_{(m)}$

\qquad{}$m=m+1$

Let $\Delta\overrightarrow{E}=\Delta\overrightarrow{E}_{(m_{\mathrm{max}})}$

\caption{Algorithm for iteration to solve Eq(\ref{eq:transform-e-e_aux})\label{alg:iteration}}
\end{algorithm}

\subsubsection{Equations for two dimensional case}

The Maxwell equations on two dimensional grid in $x-y$ plane can
be written down by setting $D_{z}\to0$ and $\alpha_{z}\to1$ in Eqs(\ref{eq:faraday_x})
to (\ref{eq:faraday_z}) and (\ref{eq:ampere_aux_x}) to (\ref{eq:ampere_aux_z}).
We obtain
\begin{align}
\frac{1}{c}D_{t}B_{x} & =-D_{y}E_{z}\label{eq:faraday_x_2d}\\
\frac{1}{c}D_{t}B_{y} & =\phantom{-}D_{x}E_{z}\\
\frac{1}{c}D_{t}B_{z} & =-D_{x}E_{y}+D_{y}E_{x}\\
\frac{1}{c}D_{t}E_{x}^{\prime} & =\phantom{-}D_{y}B_{z}-\frac{4\pi}{c}j_{x}\\
\frac{1}{c}D_{t}E_{y}^{\prime} & =-D_{x}B_{z}-\frac{4\pi}{c}j_{y}\\
\frac{1}{c}D_{t}E_{z}^{\prime} & =\phantom{-}D_{x}(\alpha_{y}B_{y})-D_{y}(\alpha_{x}B_{x})-\frac{4\pi}{c}j_{z}\label{eq:ampere_aux_z_2d}
\end{align}
The matrix elements of $\overleftrightarrow{A}$ on two dimensional
grid in $x-y$ plane can be written down by setting $\bar{D}_{z}\to0$
in Eqs(\ref{eq:matrixA_1st}) to (\ref{eq:matrixA_last}). We obtain
\begin{align}
A_{xx} & =\frac{\frac{c^{2}\Delta t^{2}}{\Delta y^{2}}-1}{6}\bar{D}_{yy}\label{eq:matrixA_1st_2d}\\
A_{yy} & =\frac{\frac{c^{2}\Delta t^{2}}{\Delta x^{2}}-1}{6}\bar{D}_{xx}\\
A_{xy} & =A_{yx}=-\frac{c^{2}\Delta t^{2}}{6\Delta x\Delta y}\bar{D}_{xy}\\
A_{zz} & =\frac{\frac{c^{2}\Delta t^{2}}{\Delta x^{2}}-1}{6}\bar{D}_{xx}+\frac{\frac{c^{2}\Delta t^{2}}{\Delta y^{2}}-1}{6}\bar{D}_{yy}\nonumber \\
 & \qquad+\frac{\frac{c^{2}\Delta t^{2}}{\Delta y^{2}}+\frac{c^{2}\Delta t^{2}}{\Delta x^{2}}-1}{36}\bar{D}_{xxyy}\\
A_{yz} & =A_{zx}=A_{zy}=A_{xz}=0\label{eq:matrixA_last_2d}
\end{align}
Using Alg \ref{alg:iteration}, $E_{x}$ and $E_{y}$ need to be solved
simultaneously because $A_{xy}\ne0$, but $E_{z}$ can be solved independent
of $E_{x}$ and $E_{y}$.

\subsubsection{Equations for one dimensional case}

The Maxwell equations on one dimensional grid in $x$ direction can
be written down by setting $D_{y}\to0$ and $\alpha_{y}\to1$ in Eqs(\ref{eq:faraday_x_2d})
to (\ref{eq:ampere_aux_z_2d}). We obtain
\begin{align}
\frac{1}{c}D_{t}B_{x} & =0\\
\frac{1}{c}D_{t}B_{y} & =\phantom{-}D_{x}E_{z}\\
\frac{1}{c}D_{t}B_{z} & =-D_{x}E_{y}\\
\frac{1}{c}D_{t}E_{x}^{\prime} & =-\frac{4\pi}{c}j_{x}\\
\frac{1}{c}D_{t}E_{y}^{\prime} & =-D_{x}B_{z}-\frac{4\pi}{c}j_{y}\\
\frac{1}{c}D_{t}E_{z}^{\prime} & =\phantom{-}D_{x}B_{y}-\frac{4\pi}{c}j_{z}
\end{align}
The matrix elements of $\overleftrightarrow{A}$ on one dimensional
grid in $x$ direction can be written down by setting $\bar{D}_{y}\to0$
in Eqs(\ref{eq:matrixA_1st_2d}) to (\ref{eq:matrixA_last_2d}). We
obtain
\begin{align}
A_{yy} & =A_{zz}=\frac{\frac{c^{2}\Delta t^{2}}{\Delta x^{2}}-1}{6}\bar{D}_{xx}\\
A_{xx} & =A_{yz}=A_{zy}=A_{xy}\nonumber \\
 & =A_{yx}=A_{xz}=A_{zx}=0
\end{align}
Using Alg \ref{alg:iteration}, $E_{y}$ and $E_{z}$ can be solved
independently and there is no need to solve for $E_{x}$ because $E_{x}=E_{x}^{\prime}$.

\section{NCI growth rate\label{sec:NCI-growth-rate}}

For a cold plasma drifting with ultra-relativistic velocity $v_{x}\to c$,
we derive in Appendix \ref{sec:Derivation-of-NCI} the NCI growth
rate following Xu 2013\citep{NCI_Xu2013}. In two dimensional grid
in $x-y$ plane, the asymptotic expression for NCI growth rate is

\begin{equation}
\Gamma(\overrightarrow{k})=\frac{\sqrt{3}}{2}\bigg|\frac{\omega_{p}^{2}c^{2}[k]_{y}k_{y}S_{J_{x}}(\alpha_{x}\xi_{0}S_{B_{z}}-\xi_{2}c[k]_{x}S_{E_{y}})}{\xi_{0}^{2}\alpha_{x}\alpha_{y}(2\xi_{1}-\xi_{3}\xi_{2}^{-1}\xi_{0})}\bigg|^{1/3}
\end{equation}
where $\omega_{p}=\sqrt{\frac{4\pi q^{2}n_{e}}{\gamma m_{e}}}$ is
the relativistic electron plasma frequency. The notation $S$ with
subscript is for the interpolation functions for the corresponding
component of $\overrightarrow{E}$, $\overrightarrow{B}$ and $\overrightarrow{J}$.
The averaging operators in Fourier space is given in Eq(\ref{eq:alphax})
and $[k]_{i}$ in Eq(\ref{eq:braket_k}). The $\xi_{0}$, $\xi_{1}$,
$\xi_{2}$, $\xi_{3}$ are defined in Eqs(\ref{eq:xi0}) to (\ref{eq:xi3}).
The growth rate is only nonzero for the condition that $(\omega^{\prime},k_{x}^{\prime},k_{y}^{\prime})$
sits near the EM modes and beam modes
\begin{align}
\alpha_{t}^{-1}\frac{[\omega]^{2}}{c^{2}} & =\alpha_{x}^{-1}[k_{x}]^{2}+\alpha_{y}^{-1}[k_{y}]^{2}\label{eq:reson1}\\
\omega^{\prime} & =ck_{x}^{\prime}\label{eq:reson2}
\end{align}

For our momentum conserving scheme we have

\begin{align}
S_{J_{x}} & =s_{l,x}s_{l,y}(-1)^{\nu_{x}}\\
S_{B_{z}} & =\cos(\frac{\omega^{\prime}\Delta t}{2})s_{l,x}s_{l,y}(-1)^{\nu_{x}}\\
S_{E_{y}} & =s_{l,x}s_{l,y}
\end{align}
where 
\begin{equation}
s_{l,i}=(\frac{\sin(k_{i}\Delta_{i}/2)}{k_{i}\Delta_{i}/2})^{l+1},\qquad l=0,1,2,\dots
\end{equation}
Thus
\begin{equation}
\Gamma(\vec{k})=\frac{\sqrt{3}}{2}\bigg|\omega_{p}s_{l,x}s_{l,y}\bigg|^{2/3}\bigg|\frac{c^{2}[k]_{y}k_{y}[\alpha_{x}\xi_{0}\cos(\frac{k_{x}^{\prime}c\Delta t}{2})(-1)^{\nu_{x}}-\xi_{2}c[k]_{x}]}{\xi_{0}^{2}\alpha_{x}\alpha_{y}(2\xi_{1}-\xi_{3}\xi_{2}^{-1}\xi_{0})}\bigg|^{1/3}\label{eq:NCI-growth-rate}
\end{equation}

\section{Numerical verifications\label{sec:Numerical-experiments}}

We implement our Maxwell solver in the PIC code EPOCH 2D\citep{EPOCH_Arber2015},
where we use the default Boris particle push and the options for particle
shapes in the code. We use $m=20$ iterations for solving Eq(\ref{eq:transform-e-e_aux}).

\subsection{2D dispersion relation for electromagnetic wave}

To initialize the simulation, we only put a point source in one cell
of the two dimensional domain, so that the spatial Fourier transform
of field in the initial frame is nonzero for all wavevectors. In the
simulations, we use $\Delta x=\Delta y$, and number of cells $N_{x}\times N_{y}=128\times128$.
The time step is $\Delta t=0.1\Delta t_{\mathrm{CFL}},0.5\Delta t_{\mathrm{CFL}},0.7\Delta t_{\mathrm{CFL}},1.0\Delta t_{\mathrm{CFL}}$.
And the simulation is run for $t_{\mathrm{max}}=256\frac{\Delta x}{c}$.
No particles are loaded in this test. In the post-processing, the
discrete Fourier transform is performed for electric field in both
space and time to test the numerical dispersion relation. To make
the data periodic in time before the discrete Fourier transform, we
apply the Hann window function 
\begin{equation}
w(t)=\sin^{2}(\frac{\pi t}{t_{\mathrm{max}}})
\end{equation}
The numerical value of the frequency for each $(k_{x},k_{y}$) is
calculated to be the weighted value of $\omega$ in the discrete Fourier
transform, i.e, 
\begin{equation}
\omega_{\mathrm{numerical}}=\frac{\sum_{\omega}\omega|\tilde{E}_{z}(\omega,k_{x,}k_{y})|^{2}}{\sum_{\omega}|\tilde{E}_{z}(\omega,k_{x,}k_{y})|^{2}}
\end{equation}
where $\tilde{E}_{z}(\omega,k_{x,}k_{y})$ is the discrete Fourier
transform of $E_{z}(t,x,y)$, $\omega$ is the discretized frequency
in the discrete Fourier transform, and $(k_{x},k_{y})$ is the discretized
wavevector. Since $E_{z}(t,x,y)$ is always real-valued, the transform
in $t\to\omega$ satisfies $\tilde{E}_{z}(\omega)=\tilde{E}_{z}^{*}(-\omega)$,
thus it is sufficient to calculate the $\omega\ge0$ components. The
results for phase velocities in the numerical tests are plotted in
Fig. \ref{fig:dispersion-EM2d} (d) to (f) against the phase velocities
calculated from Eq(\ref{eq:dispersion-em-F-2d}). The results of numerical
experiments are consistent with Eq(\ref{eq:dispersion-em-F-2d}) for
$\Delta t=0.5\Delta t_{\mathrm{CFL}},0.7\Delta t_{\mathrm{CFL}},1.0\Delta t_{\mathrm{CFL}}$,
which suggests that the number of iterations is sufficient.  The EM
wave at high $\overrightarrow{k}$ becomes more superluminal, i.e.
larger $v_{ph}$, for smaller time steps.

\subsection{Location of NCI in $\protect\overrightarrow{k}$ space and growth
rate}

\begin{table*}
\caption{Parameters for test problem of a drifting pair plasma. The definition
of electron skin depth is $d_{e}=\frac{c}{\omega_{p}}=c/\sqrt{\frac{4\pi q^{2}n_{e}}{\gamma m_{e}}}$.\label{tab:Parameters-for-test}}

\begin{tabular}{|c|c|}
\hline 
domain size & $L_{x}=16d_{e}$, $L_{y}=8d_{e}$\tabularnewline
\hline 
boundary condition & periodic, both in $x$ and $y$\tabularnewline
\hline 
number of cells & $N_{x}=256$, $N_{y}=128$\tabularnewline
\hline 
pseudo-particles per cell & $N_{\mathrm{PPC}}=512$ (256 for each species)\tabularnewline
\hline 
drift Lorentz factor & $\gamma=1000$\tabularnewline
\hline 
temperature & $k_{B}T_{e}=k_{B}T_{i}=0.01m_{e}c^{2}$\tabularnewline
\hline 
time step & $\Delta t/\Delta t_{\mathrm{CFL}}=0.1,0.2,\dots1.0$\tabularnewline
\hline 
particle shape & 1th, 2nd, 4th order B-spline\tabularnewline
\hline 
\end{tabular}
\end{table*}

We carry out numerical tests to verify the location of NCI in $\overrightarrow{k}$
space and compare the growth rate in the test runs with the analytical
expression in Eq(\ref{eq:NCI-growth-rate}). In our test runs, a pair
plasma is initialize with a drifting speed $\beta$ in $x$ direction.
And the Lorentz factor is $\gamma=\frac{1}{\sqrt{1-\beta^{2}}}$.
The numerical parameters of the test runs are listed in Table \ref{tab:Parameters-for-test}.
We use the method in \citep{Zenitani2015} to load the particles with
relativistic shifted-Maxwellian distribution, which takes care of
the spatial part of the transformation $d^{3}x\to d^{3}x^{\prime}$,
and the acceptance efficiencies are 50\% for generic cases and 100\%
for symmetric distributions. We use momentum conserving scheme for
field interpolation and current deposition in our simulations.

\begin{figure*}
\includegraphics[scale=0.3]{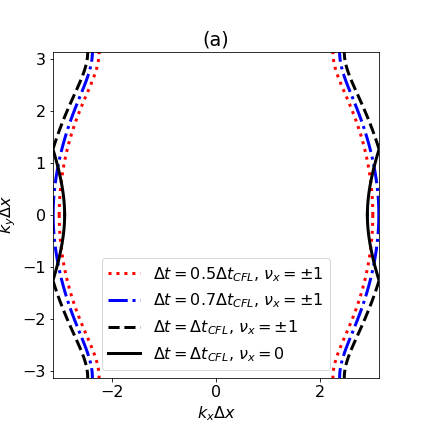}\includegraphics[scale=0.3]{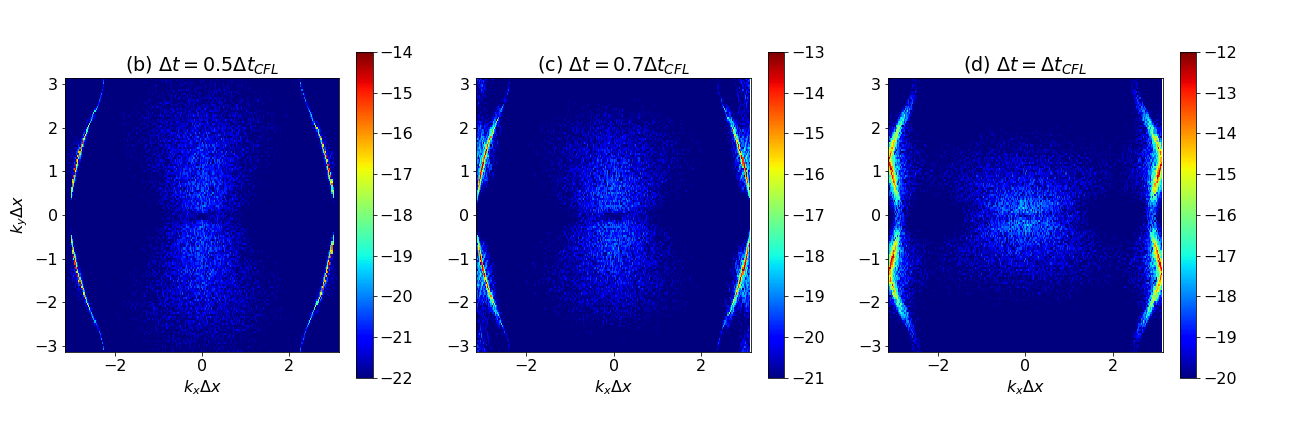}

\includegraphics[scale=0.3]{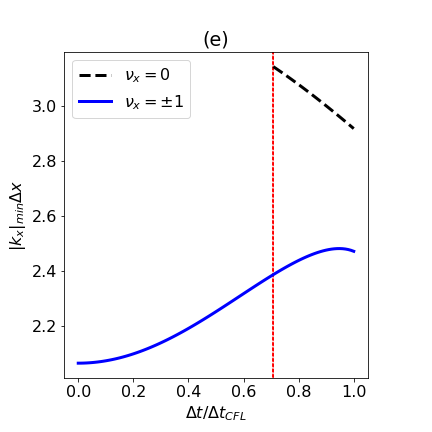}\includegraphics[scale=0.3]{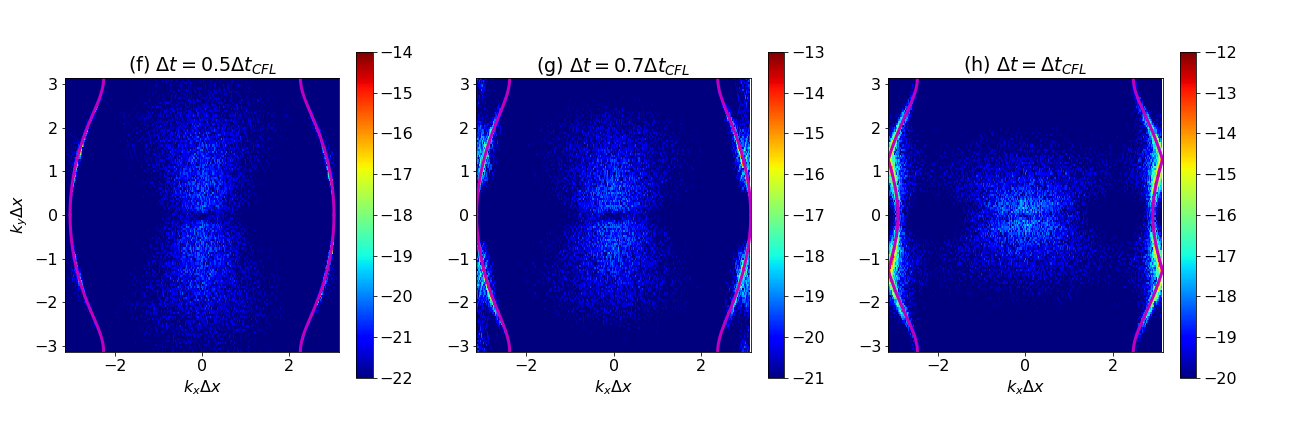}

\includegraphics[scale=0.5]{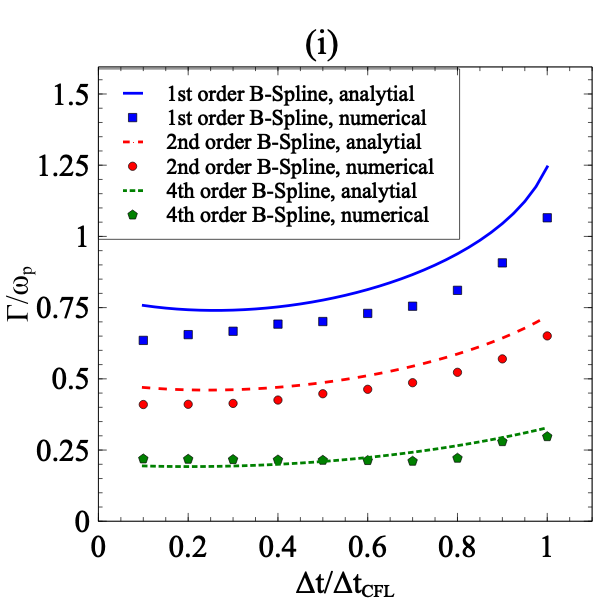}

\caption{Theoretical and numerical tests for the properties of NCI using our
Maxwell solver. Subfigure (a) shows the location of NCI by solving
Eq(\ref{eq:reson1}) and (\ref{eq:reson2}) for the main beam $\nu_{x}=0$
and first order aliasing beam $\nu_{x}=\pm1$, there is no higher
order aliasing $|\nu_{x}|\ge2$ as solved by using Eq(\ref{eq:reson1})
and (\ref{eq:reson2}). Subfigures (b) to (d) shows the color-coded
spatial Fourier transform of out-of-plane component of the magnetic
field $B_{z}$ in the simulation on log scale, the unit is arbitrary
but same among subfigures (b) to (d). It can be seen that the NCI
grows to a higher level when the time step is increased and the aliased
beam resonance dominates the main mean resonance. Subfigures (f) to
(h) are same as (b) to (d) but overlaid with the theoretical curve
for $(k_{x},k_{y})$ in subfigure (a) by mega curve. Subfigure (e)
shows the minimum value of $|k_{x}|$ in the NCI modes vs. time step,
for $\nu_{x}=0,\pm1$. Subfigure (i) shows the maximum NCI growth
rate calculated from EPOCH simulations vs. the analytical formula
Eq(\ref{eq:NCI-growth-rate}) for different order of particle shape
functions and different time step $\Delta t$. The scattered points
are the half of the growth rate of electromagnetic energy, while the
curves are the growth rate of the amplitude of the most unstable mode
calculated from analytical formula Eq(\ref{eq:NCI-growth-rate}).\label{fig:NCI-vis}}
\end{figure*}

The locations of NCI resonance modes in $\overrightarrow{k}$ space
are computed by solving Eq(\ref{eq:reson1}) and (\ref{eq:reson2})
for the main beam $\nu_{x}=0$ and first order aliasing beam $\nu_{x}=\pm1$,
as shown in Fig. \ref{fig:NCI-vis} (a), where $\nu_{x}$ is the index
for aliasing beam as in Eq(\ref{eq:aliasing_x}). The smallest value
of $|k_{x}|$ in the NCI modes is plotted in Fig. \ref{fig:NCI-vis}(e),
showing that if $\Delta t<\frac{\Delta t_{\mathrm{CFL}}}{\sqrt{2}}$
only $\nu_{x}=\pm1$ resonance exists, if $\frac{\Delta t_{\mathrm{CFL}}}{\sqrt{2}}<\Delta t<\Delta t_{\mathrm{CFL}}$
both $\nu_{x}=0$ and $\nu_{x}=\pm1$ resonance exist. However, the
simulation shows that the dominant NCI mode is from the $\nu_{x}=\pm1$
resonance which indicates that an effective method to suppress such
resonance is desired. The location of NCI modes is bounded by $\frac{2.06}{\Delta x}<|k_{x}|<\frac{\pi}{\Delta x}$
for all the valid time step $\Delta t\le\Delta t_{\mathrm{CFL}}$.
The NCI modes in the numerical experiments are visualized in Fig.
\ref{fig:NCI-vis} (b)-(d) by plotting the spatial Fourier transform
of the out-of-plane magnetic field. The locations of NCI in $\overrightarrow{k}$
space are consistent between the theory and the numerical experiments
as shown in Fig. \ref{fig:NCI-vis} (f)-(h). The growth rate in Eq(\ref{eq:NCI-growth-rate})
is essentially zero on $k_{x}$ axis, which is also seen in Fig. \ref{fig:NCI-vis}
(b)-(d). There is no higher order aliasing $|\nu_{x}|\ge2$, as solved
by using Eq(\ref{eq:reson1}) and (\ref{eq:reson2}).

In the test runs for drifting pair plasma, the fastest growing NCI
mode will dominate in the growth of the total field energy, thus the
growth rate for the amplitude of the fastest growing NCI mode is half
of the growth rate for the total field energy for WKB approximation.
The growth rate of the fastest growing NCI mode can be calculated
by maximizing Eq(\ref{eq:NCI-growth-rate}) over $k_{x}$ or $k_{y}$.
We calculate the growth rate of the fastest growing NCI mode, i.e.
the most unstable NCI mode, as shown in the curves in Fig. \ref{fig:NCI-vis}(i).
In the same figure, we also show the growth rate calculated from the
numerical experiments by calculating half of the growth rate of total
field energy. Using high order B-spline particle shape significantly
reduces the NCI growth rate, although more computational cost is required.
The trend of the NCI growth rate are consistent between the analytical
expression and the numerical results.

\begin{figure*}
\includegraphics[scale=0.33]{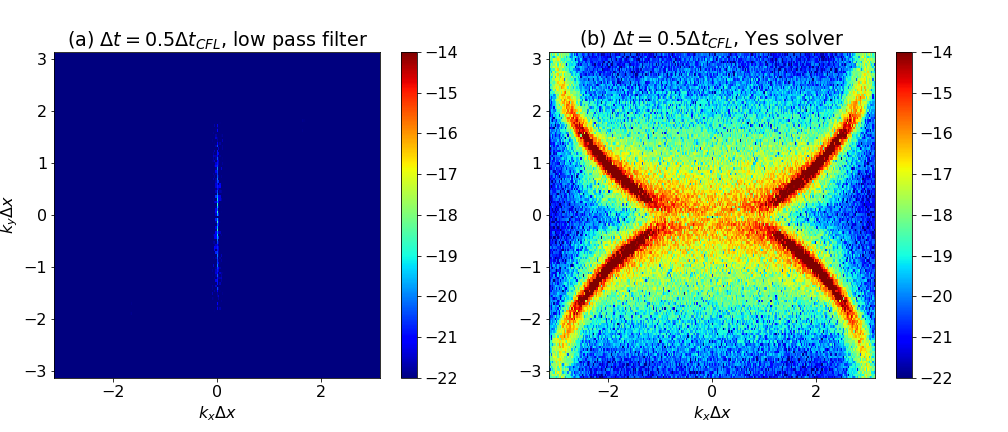}\includegraphics[scale=0.4]{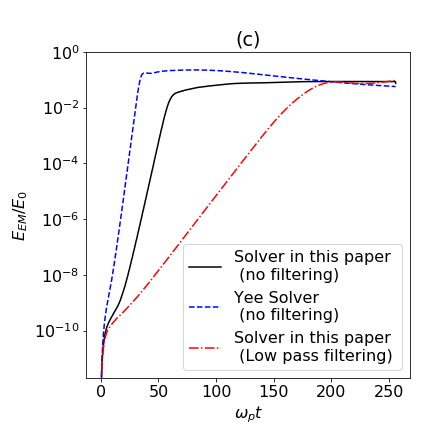}

\caption{Results for numerical tests using low pass filtering and comparison
to Yee solver, at $\omega_{p}t=50$. (a) Color-coded spatial Fourier
transform of out-of-plane component of magnetic field $B_{z}$ in
the simulation on log scale in the run with low pass filtering and
$\Delta t=0.5\Delta t_{\mathrm{CFL}}$, the unit and color scale is
as same as Fig. \ref{fig:NCI-vis}(b). It can be seen that the NCI
is significantly suppressed. (b) Same as (a) but with Yee solver and
no low pass filtering. (c) The growth curve for total electromagnetic
energy $E_{EM}$ (normalized by initial total particle energy $E_{0}$).
The growth rate of $E_{EM}$ is $0.71\text{\ensuremath{\omega_{p}}}$
for Yee solver, $0.41\text{\ensuremath{\omega_{p}}}$ for the solver
in this paper without filtering, and $0.12\text{\ensuremath{\omega_{p}}}$
for the solver in this paper with low pass filtering. \label{fig:NCI-vis-improve}}
\end{figure*}

For the $\Delta t=0.5\Delta t_{\mathrm{CFL}}$ case, we use the low
pass filtering with our solver and make the comparison with Yee solver,
as shown in Fig. \ref{fig:NCI-vis-improve}. The growth rate of NCI
is significantly suppressed as shown in Fig. \ref{fig:NCI-vis-improve}(a)
and (c).

\section{Conclusions and discussions}

We demonstrate that the Maxwell solver from Eastwood\citep{Eastwood1991}
has several good numerical properties that allow for the efficient
mitigation of NCI. The dispersion relation is always superluminal
in a large range of $k$ including $k=0$, thus the NCI resonance
are separated from the physical modes in $k$ space. There is only
high $k$ NCI mode, i.e. $\frac{2.06}{\Delta x}<|k_{x}|<\frac{\pi}{\Delta x}$
for $\nu_{x}=0$ and $\nu_{x}=\pm1$ NCI. Those high $k$ modes can
be suppressed by applying a low-pass filter on the current during
the simulation. An alternative approach to deal with the high $k$
mode is to post-process the fields and currents with a low-pass filter
after the simulation. Although our algorithm requires solving linear
equations using an iteration method, it is local, of $\mathcal{O}(N)$
computational cost and scalable for multi-dimensional simulation applications,
which has advantage over the methods based on spatial Fourier transform.
This solver also possess conservation properties such that no correction
scheme is needed. The examples in this paper are using 2D grid, but
our implementation and discussions on the numerical properties can
be generalized to 3D grid. The averaging stencil we used in this work
is $(\frac{1}{6},\frac{2}{3},\frac{1}{6})$. Another stencil can be
potentially found to further improve the numerical properties. The
rate of convergence may also be improved by using more optimized iteration
scheme. These further improvements and generalizations will be subjects
of future reports.

The Maxwell solver we developed was derived from finite element method
in Eastwood 1991\citep{Eastwood1991}. The particle pusher we use
is the Boris push, which conserves the phase space volume for particles.
If we derive the particle pusher using the finite element method in
Eastwood 1991, then the particle pusher should be implicit. By the
combination of that implicit particle pusher and the Maxwell solver
we used in this work, the numerical solution of all the particle and
field variables should strictly satisfy the minimal action principle
and has better conservation properties than traditional PIC methods.
As we learned from Eastwood 1991, the numerical conservation laws
can be well defined and without error terms in the finite element
method. With finite element methods, we can potentially construct
numerical schemes which is structure preserving for both particles
and fields in the relativistic case, as an alternative approach to
the method derived from Hamiltonian for the non-relativistic case\citep{symplectic_Qin2015}.
Further more, the finite element PIC is suitable for complex geometries.

\section{Acknowledgements}

Research presented in this paper was supported by the Center for Space
and Earth Science(CSES) program and Laboratory Directed Research and
Development(LDRD) program of Los Alamos National Laboratory(LANL).
The simulations were performed with LANL Institutional Computing which
is supported by the U.S. Department of Energy National Nuclear Security
Administration under Contract No. 89233218CNA000001, and with the
Extreme Science and Engineering Discovery Environment (XSEDE), which
is supported by National Science Foundation(NSF) grant number ACI-1548562.

\appendix

\section{Conservation of the divergence of the fields\label{sec:Conservation}}

Starting from the L.H.S of Eq(\ref{eq:gauss_B}) and apply $D_{t}$
on it, and using Eqs(\ref{eq:faraday_x}) to (\ref{eq:faraday_z}),
we obtain
\begin{align}
 & \frac{1}{c}D_{t}(D_{x}B_{x}+D_{y}B_{y}+D_{z}B_{z})\nonumber \\
= & \frac{1}{c}\bigg[D_{x}(D_{t}B_{x})+D_{y}(D_{t}B_{y})+D_{z}(D_{t}B_{z})\bigg]\nonumber \\
= & D_{x}(-D_{y}E_{z}+D_{z}E_{y})\nonumber \\
 & \qquad+D_{y}(-D_{z}E_{x}+D_{x}E_{z})\nonumber \\
 & \qquad+D_{z}(-D_{x}E_{y}+D_{y}E_{x})\nonumber \\
= & (D_{z}D_{y}-D_{y}D_{z})E_{x}+(D_{x}D_{z}-D_{z}D_{x})E_{y}\nonumber \\
 & \qquad+(D_{y}D_{x}-D_{x}D_{y})E_{z}\nonumber \\
= & 0
\end{align}
Thus Eq(\ref{eq:gauss_B}) is conserved during the evolution as long
as it is conserved initially.

The time evolution of the difference between the L.H.S of Eq(\ref{eq:gauss_E})
can be calculated by using Eqs(\ref{eq:ampere_x}) to (\ref{eq:ampere_z})
and the continuity equation for charge conserving deposition scheme
\begin{equation}
D_{t}\rho+D_{x}j_{x}+D_{y}j_{y}+D_{z}j_{z}=0
\end{equation}
We obtain
\begin{align}
 & \frac{1}{c}\bigg[D_{t}\big(D_{x}(\alpha_{yz}E_{x})+D_{y}(\alpha_{zx}E_{y})+D_{z}(\alpha_{xy}E_{z})\big)\bigg]\nonumber \\
= & \frac{1}{c}\bigg[D_{x}\big(D_{t}(\alpha_{yz}E_{x})\big)+D_{y}\big(D_{t}(\alpha_{zx}E_{y})\big)+D_{z}\big(D_{t}(\alpha_{xy}E_{z})\big)\bigg]\nonumber \\
= & D_{x}\big(D_{y}(\alpha_{tz}B_{z})-D_{z}(\alpha_{ty}B_{y})-\frac{4\pi}{c}j_{x}\big)\nonumber \\
 & \qquad+D_{y}\big(D_{z}(\alpha_{tx}B_{x})-D_{x}(\alpha_{tz}B_{z})-\frac{4\pi}{c}j_{y}\big)\nonumber \\
 & \qquad+D_{z}\big(D_{x}(\alpha_{ty}B_{y})-D_{y}(\alpha_{tx}B_{x})-\frac{4\pi}{c}j_{z}\big)\nonumber \\
= & (D_{y}D_{z}-D_{z}D_{y})(\alpha_{tx}B_{x})+(D_{z}D_{x}-D_{x}D_{z})(\alpha_{ty}B_{y})\nonumber \\
 & \qquad+(D_{x}D_{y}-D_{y}D_{x})(\alpha_{tz}B_{x})\nonumber \\
 & \qquad-\frac{4\pi}{c}(D_{x}j_{x}+D_{y}j_{z}+D_{z}j_{z})\nonumber \\
= & \frac{4\pi}{c}D_{t}\rho
\end{align}
where the last step requires the use of a charge-conserving current
deposition scheme. Thus Eq(\ref{eq:gauss_E}) is conserved during
the evolution as long as it is conserved initially and the charge
conserving deposition scheme is used. If one does not use a charge-conserving
current deposition scheme, then the correction for the the Gauss's
law for electric field is needed, otherwise the finite grid instability
can grow due to the loss of charge conservation\citep{ES_Meyers2015}.

\section{Derivation for the limit of time step and the superluminal property
of dispersion relation \label{sec:Derivation-inequties}}

For real value of $\omega$, the function $G(\frac{\omega\Delta t}{2})$
satisfies $0\le G(\frac{\omega\Delta t}{2})\le3.$ From Eq(\ref{eq:dispersion-em-F-3d})
we need for real $\omega$ that 
\begin{equation}
c^{2}\Delta t^{2}\bigg[\frac{G(\frac{k_{x}\Delta x}{2})}{\Delta x^{2}}+\frac{G(\frac{k_{y}\Delta y}{2})}{\Delta y^{2}}+\frac{G(\frac{k_{z}\Delta z}{2})}{\Delta z^{2}}\bigg]\le3
\end{equation}
So
\begin{equation}
c\Delta t\le\sqrt{\frac{3}{\frac{G(\frac{k_{x}\Delta x}{2})}{\Delta x^{2}}+\frac{G(\frac{k_{y}\Delta y}{2})}{\Delta y^{2}}+\frac{G(\frac{k_{z}\Delta z}{2})}{\Delta z^{2}}}}
\end{equation}
And $\Delta t$ needs to be less or equal to the minimum value on
the R.H.S, which is reached when $G(\frac{k_{y}\Delta x}{2})=G(\frac{k_{y}\Delta x}{2})=G(\frac{k_{y}\Delta x}{2})=3$,
thus we get Eq(\ref{eq:dt_cfl_3d})
\begin{equation}
c\Delta t\le c\Delta t_{\mathrm{3D,CFL}}=(1/\Delta x^{2}+1/\Delta y^{2}+1/\Delta z^{2})^{-1/2}
\end{equation}
The 2D case can be derived in the same way.

From (\ref{eq:phase-speed-3d}) we have
\begin{align}
\omega/k & =c\bigg[1+\frac{c^{2}\Delta t^{2}}{24k^{2}}(\frac{k_{x}^{4}\Delta x^{2}+k_{y}^{4}\Delta y^{2}+k_{z}\Delta z^{2}}{c^{2}\Delta t^{2}}-k^{4})\nonumber \\
 & \phantom{=c\bigg[1}+\mathcal{O}(k^{4})
\end{align}
Using Eq(\ref{eq:dt_cfl_3d}) we have
\begin{align}
 & \frac{k_{x}^{4}\Delta x^{2}+k_{y}^{4}\Delta y^{2}+k_{z}\Delta z^{2}}{c^{2}\text{\ensuremath{\Delta t^{2}}}}-k^{4}\nonumber \\
\ge & (k_{x}^{4}\Delta x^{2}+k_{y}^{4}\Delta y^{2}+k_{z}\Delta z^{2})\nonumber \\
 & \qquad\times(1/\Delta x^{2}+1/\Delta y^{2}+1/\Delta z^{2})\nonumber \\
 & \qquad-(k_{x}^{2}+k_{y}^{2}+k_{z}^{2})^{2}\label{eq:proof-superlum}
\end{align}
Using Cauchy inequality $(a_{1}^{2}+a_{2}^{2}+a_{3}^{2})(b_{1}^{2}+b_{2}^{2}+b_{3}^{2})\ge(a_{1}b_{1}+a_{2}b_{2}+a_{3}b_{3})^{2}$
and letting $a_{1}=k_{x}^{2}\Delta x$, $a_{2}=k_{y}^{2}\Delta y$,
$a_{3}=k_{z}^{2}\Delta z$, $b_{1}=\frac{1}{\Delta x}$, $b_{2}=\frac{1}{\Delta y}$,
$b_{3}=\frac{1}{\Delta z}$, then the R.H.S of Eq(\ref{eq:proof-superlum})
is greater than or equal to zero, thus we have $\omega/k\ge c$. For
the L.H.S. of Eq(\ref{eq:proof-superlum}) to be equal to zero, we
need $c\Delta t=\frac{1}{(1/\Delta x^{2}+1/\Delta y^{2}+1/\Delta z^{2})^{1/2}}$
and $\frac{a_{1}}{b_{1}}=\frac{a_{2}}{b_{2}}=\frac{a_{3}}{b_{3}}$,
i.e. $|k_{x}|\Delta x|=|k_{y}|\Delta y=|k_{z}|\Delta z$. The 2D case
can be derived in the same way.

\section{Derivation of the inequality for infinite norm of matrix $\protect\overleftrightarrow{A}$
\label{sec:Derivation-of-norm}}

The infinite norm of matrix $\overleftrightarrow{A}$ is simply the
maximum absolute row sum of the matrix. The absolute $x$ row sum
is $\sum_{j=x,y,z}|A_{xj}|=|A_{xx}|+|A_{xy}|+|A_{xz}|$. The operators
$\bar{D}_{x},\bar{D}_{y}$, etc are diagonalized in Fourier space,
thus it is convenient to analyze the properties of matrix $\overleftrightarrow{A}$
in Fourier space. In Fourier space $\bar{D}_{x}=2i\sin(k_{x}\Delta x/2)$
satisfies $|\bar{D}_{x}|\le2$. Similarly, $|\bar{D}_{y}|\le2$ and
$|\bar{D}_{z}|\le2$. And $\bar{D}_{xx}=-4\sin^{2}(k_{x}\Delta x/2)$
satisfy $-4\le\bar{D}_{xx}\le0$. Similarly, $-4\le\bar{D}_{yy}\le0$
and $-4\le\bar{D}_{zz}\le0$. Thus
\begin{align}
A_{xx} & =\frac{1-\frac{c^{2}\Delta t^{2}}{\Delta y^{2}}}{6}\bar{D}_{yy}+\frac{1-\frac{c^{2}\Delta t^{2}}{\Delta z^{2}}}{6}\bar{D}_{zz}\nonumber \\
 & \qquad+\frac{1-\frac{c^{2}\Delta t^{2}}{\Delta y^{2}}-\frac{c^{2}\Delta t^{2}}{\Delta z^{2}}}{36}\bar{D}{}_{yyzz}\nonumber \\
 & =\frac{1-\frac{c^{2}\Delta t^{2}}{\Delta y^{2}}-\frac{c^{2}\Delta t^{2}}{\Delta z^{2}}}{36}\bigg(\bar{D}_{yy}+\frac{6(1-\frac{c^{2}\Delta t^{2}}{\Delta z^{2}})}{1-\frac{c^{2}\Delta t^{2}}{\Delta y^{2}}-\frac{c^{2}\Delta t^{2}}{\Delta z^{2}}}\bigg)\nonumber \\
 & \qquad\times\bigg(\bar{D}_{zz}+\frac{6(1-\frac{c^{2}\Delta t^{2}}{\Delta y^{2}})}{1-\frac{c^{2}\Delta t^{2}}{\Delta y^{2}}-\frac{c^{2}\Delta t^{2}}{\Delta z^{2}}}\bigg)\nonumber \\
 & \qquad-\frac{(1-\frac{c^{2}\Delta t^{2}}{\Delta y^{2}})(1-\frac{c^{2}\Delta t^{2}}{\Delta y^{2}})}{1-\frac{c^{2}\Delta t^{2}}{\Delta y^{2}}-\frac{c^{2}\Delta t^{2}}{\Delta z^{2}}}
\end{align}
We use Eq(\ref{eq:dt_cfl_3d}), which makes $1-\frac{c^{2}\Delta t^{2}}{\Delta y^{2}}>0$,
$1-\frac{c^{2}\Delta t^{2}}{\Delta z^{2}}>0$, $1-\frac{c^{2}\Delta t^{2}}{\Delta y^{2}}-\frac{c^{2}\Delta t^{2}}{\Delta z^{2}}>0$
and $\frac{1-\frac{c^{2}\Delta t^{2}}{\Delta y^{2}}}{1-\frac{c^{2}\Delta t^{2}}{\Delta y^{2}}-\frac{c^{2}\Delta t^{2}}{\Delta z^{2}}}>1.$
The maximum of $A_{xx}$ is 0 when $\bar{D}_{yy}=\bar{D}_{zz}=0$,
thus $|A_{xx}|=-A_{xx}$. Using $\bigg|\bar{D}_{zx}\bigg|\le4$, $\bigg|\bar{D}_{xy}\bigg|\le4$
we have

\begin{align}
|A_{xy}| & =\frac{c^{2}\Delta t^{2}}{6\Delta z\Delta x}\bigg|\bar{D}_{zx}\bigg|\bigg|1+\frac{\bar{D}_{yy}}{6}\bigg|\nonumber \\
 & \le\frac{2c^{2}\Delta t^{2}}{3\Delta z\Delta x}(1+\frac{\bar{D}_{yy}}{6})\\
|A_{xz}| & =\frac{c^{2}\Delta t^{2}}{6\Delta x\Delta y}\bigg|\bar{D}_{xy}\bigg|\bigg|1+\frac{\bar{D}_{zz}}{6}\bigg|\nonumber \\
 & \le\frac{2c^{2}\Delta t^{2}}{3\Delta x\Delta y}(1+\frac{\bar{D}_{zz}}{6})
\end{align}
Thus for the absolute $x$ row sum
\begin{align}
\sum_{j=x,y,z}|A_{xj}| & \le-\frac{1-\frac{c^{2}\Delta t^{2}}{\Delta y^{2}}}{6}\bar{D}_{yy}-\frac{1-\frac{c^{2}\Delta t^{2}}{\Delta z^{2}}}{6}\bar{D}_{zz}\nonumber \\
 & \qquad-\frac{1-\frac{c^{2}\Delta t^{2}}{\Delta y^{2}}-\frac{c^{2}\Delta t^{2}}{\Delta z^{2}}}{36}\bar{D}{}_{yyzz}\nonumber \\
 & \qquad+\frac{2c^{2}\Delta t^{2}}{3\Delta z\Delta x}(1+\frac{\bar{D}_{yy}}{6})\nonumber \\
 & \qquad+\frac{2c^{2}\Delta t^{2}}{3\Delta x\Delta y}(1+\frac{\bar{D}_{zz}}{6})\label{eq:inf-norm-A-3D}
\end{align}
The R.H.S. of Eq(\ref{eq:inf-norm-A-3D}) is monotonic in $\bar{D}_{yy}$
and $\bar{D}_{zz}$ as derived in Appendix \ref{subsec:Derivation-of-the-inequality},
and reaches maximum when $\bar{D}_{yy}=\bar{D}_{zz}=-2$ , thus
\begin{align}
\sum_{j=x,y,z}|A_{xj}| & \le\frac{8}{9}-\frac{2}{9}\frac{c^{2}\Delta t^{2}}{\Delta y^{2}}-\frac{2}{9}\frac{c^{2}\Delta t^{2}}{\Delta z^{2}}+\frac{2c^{2}\Delta t^{2}}{9\Delta z\Delta x}+\frac{2c^{2}\Delta t^{2}}{9\Delta x\Delta y}\nonumber \\
 & \le\frac{8}{9}-\frac{1}{9}\frac{c^{2}\Delta t^{2}}{\Delta y^{2}}-\frac{1}{9}\frac{c^{2}\Delta t^{2}}{\Delta z^{2}}+\frac{2c^{2}\Delta t^{2}}{9\Delta x^{2}}\nonumber \\
 & <\frac{8}{9}+\frac{2c^{2}\Delta t^{2}}{9\Delta x^{2}}\label{eq:inf-norm-A-3D_derive}
\end{align}
When $\Delta t\le\frac{\Delta x}{\sqrt{2}c}$ we have $\sum_{j=x,y,z}|A_{xj}|<1$.
Similarly, we have $\sum_{j=x,y,z}|A_{yj}|<1$ if $\Delta t\le\frac{\Delta y}{\sqrt{2}c}$
and we have $\sum_{j=x,y,z}|A_{zj}|<1$ if $\Delta t\le\frac{\Delta z}{\sqrt{2}c}$.
Thus the infinite norm of matrix $A$ is less than $1$ if $\Delta t\le\frac{\min(\Delta x,\Delta y,\Delta z)}{\sqrt{2}c}$.
In the $\Delta x=\Delta y=\Delta z$ case the infinite norm of matrix
$A$ is always less than $1$ because $\Delta t\le\frac{\Delta x}{\sqrt{3}c}$.

For 2D case, from Eqs(\ref{eq:matrixA_1st_2d}) to (\ref{eq:matrixA_last_2d})
we have
\begin{align}
\sum_{j=x,y}|A_{xj}| & \le\frac{2}{3}(1-\frac{c^{2}\Delta t^{2}}{\Delta y^{2}})+\frac{2c^{2}\Delta t^{2}}{3\Delta x\Delta y}\nonumber \\
 & \le\frac{2}{3}(1-\frac{c^{2}\Delta t^{2}}{\Delta y^{2}})+\frac{c^{2}\Delta t^{2}}{3\Delta x^{2}}+\frac{c^{2}\Delta t^{2}}{3\Delta y^{2}}\nonumber \\
 & <\frac{2}{3}+\frac{c^{2}\Delta t^{2}}{3\Delta x^{2}}<1
\end{align}
Similarly we can prove $\sum_{j=x,y}|A_{yj}|<1$. And
\begin{equation}
|A_{zz}|\le\frac{8}{9}-\frac{2}{9}\frac{c^{2}\Delta t^{2}}{\Delta x^{2}}-\frac{2}{9}\frac{c^{2}\Delta t^{2}}{\Delta y^{2}}<1
\end{equation}

For 1D case we have 
\begin{equation}
|A_{yy}|=|A_{zz}|\le\frac{2}{3}(1-\frac{c^{2}\Delta t^{2}}{\Delta y^{2}})<1
\end{equation}

To calculate the speed of convergence of the iteration, we need to
compute the eigenvalues of matrix $\overleftrightarrow{A}$. We derive
in Appendix \ref{sec:Speed-of-convergence} how the error of the dispersion
relation depends on the number of iterations. For symmetric matrix
$\overleftrightarrow{A}$, the absolute value of eigenvalue $|\lambda|$
is always smaller or equal to the infinity norm, thus $|\lambda|<1$
and $(\mathbb{1}-\overleftrightarrow{A})$ is invertible if the infinity
norm is less than 1.

\subsection{Derivation of the inequality for the infinite norm of $\protect\overleftrightarrow{A}$
in 3D\label{subsec:Derivation-of-the-inequality}}

The R.H.S of Eq(\ref{eq:inf-norm-A-3D}) can be written as 
\begin{equation}
f=a_{y1}d_{y}+a_{z1}d_{z}-a_{yz}d_{y}d_{z}+a_{y2}(1-d_{y})+a_{z2}(1-d_{z})
\end{equation}
where the constants $a_{y1}$, $a_{z1}$, $a_{xy}$, $a_{y2}$, $a_{z2}$
are 
\begin{align}
a_{y1} & =1-\frac{c^{2}\Delta t^{2}}{\Delta y^{2}},\qquad a_{z1}=1-\frac{c^{2}\Delta t^{2}}{\Delta z^{2}}\nonumber \\
a_{yz} & =1-\frac{c^{2}\Delta t^{2}}{\Delta y^{2}}-\frac{c^{2}\Delta t^{2}}{\Delta z^{2}}\nonumber \\
a_{y2} & =\frac{2c^{2}\Delta t^{2}}{3\Delta z\Delta x},\qquad d_{z2}=\frac{2c^{2}\Delta t^{2}}{3\Delta x\Delta y}
\end{align}
and the variables $d_{y}=-\bar{D}_{yy}/6$ and $d_{z}=-\bar{D}_{zz}/6$
satisfy $0\le d_{y}\le\frac{2}{3}$ and $0\le d_{z}\le\frac{2}{3}$.
By take the derivative of $f$ w.r.t $d_{y}$ and $d_{z}$ we can
check the monotonicity, e.g. 
\begin{align}
\frac{\partial f}{\partial d_{y}} & =a_{y1}-\alpha_{yz}d_{z}-\alpha_{y2}\nonumber \\
 & =1-\frac{c^{2}\Delta t^{2}}{\Delta y^{2}}-\big(1-\frac{c^{2}\Delta t^{2}}{\Delta y^{2}}-\frac{c^{2}\Delta t^{2}}{\Delta z^{2}}\big)d_{z}-\frac{2c^{2}\Delta t^{2}}{3\Delta z\Delta x}
\end{align}
Since $\big(1-\frac{c^{2}\Delta t^{2}}{\Delta y^{2}}-\frac{c^{2}\Delta t^{2}}{\Delta z^{2}}\big)>0$,
$\frac{\partial f}{\partial d_{y}}$ reaches minimum at $d_{z}=\frac{2}{3}$
\begin{align}
\frac{\partial f}{\partial d_{y}}\big|_{d_{z}=\frac{2}{3}} & =1-\frac{c^{2}\Delta t^{2}}{\Delta y^{2}}-\frac{2}{3}\big(1-\frac{c^{2}\Delta t^{2}}{\Delta y^{2}}-\frac{c^{2}\Delta t^{2}}{\Delta z^{2}}\big)-\frac{2c^{2}\Delta t^{2}}{3\Delta z\Delta x}\nonumber \\
 & =\frac{1}{3}-\frac{c^{2}\Delta t^{2}}{3\Delta y^{2}}+\frac{2c^{2}\Delta t^{2}}{3\Delta z^{2}}-\frac{2c^{2}\Delta t^{2}}{3\Delta z\Delta x}\nonumber \\
 & \ge\frac{1}{3}-\frac{c^{2}\Delta t^{2}}{3\Delta y^{2}}+\frac{2c^{2}\Delta t^{2}}{3\Delta z^{2}}-\frac{c^{2}\Delta t^{2}}{3\Delta z^{2}}-\frac{c^{2}\Delta t^{2}}{3\Delta x^{2}}\nonumber \\
 & =\frac{1}{3}\big(1-\frac{c^{2}\Delta t^{2}}{\Delta x^{2}}-\frac{c^{2}\Delta t^{2}}{\Delta y^{2}}-\frac{c^{2}\Delta t^{2}}{\Delta z^{2}}\big)+\frac{2c^{2}\Delta t^{2}}{3\Delta z^{2}}\nonumber \\
 & >0
\end{align}
Thus $f$ increases as $d_{y}$ increases. Similarly $f$ increases
as $d_{z}$ increases.

\section{Speed of convergence and error term for iteration\label{sec:Speed-of-convergence}}

We constrain the derivation to the case where the infinity norm of
matrix $\overleftrightarrow{A}$ is less than 1, which is always true
in 1D or 2D. Using finite number of iterations for Alg \ref{alg:iteration},
we have the error term, which is the difference between the increment
electric field we get after $m$ iterations and after infinite number
of iterations
\begin{align}
\Delta\overrightarrow{E}_{(m)}-\Delta\overrightarrow{E}_{(\infty)} & =\Delta\overrightarrow{E}^{\prime}+\overleftrightarrow{A}\cdot\Delta\overrightarrow{E}_{(m-1)}-\Delta\overrightarrow{E}_{(\infty)}\nonumber \\
 & =\overleftrightarrow{A}\cdot(\Delta\overrightarrow{E}_{(m-1)}-\Delta\overrightarrow{E}_{(\infty)})\label{eq:derive_convergence_0}
\end{align}
where we used in the last step
\begin{equation}
(\mathbb{1}-\overleftrightarrow{A})\cdot\Delta\overrightarrow{E}_{(\infty)}=\Delta\overrightarrow{E}^{\prime}\label{eq:derive_convergence_1}
\end{equation}
which is true for the exact solution $\Delta\overrightarrow{E}_{(\infty)}$
in Eq(\ref{eq:transform-e-e_aux}). Thus by mathematical induction
from Eq(\ref{eq:derive_convergence_0}) we have 
\begin{align}
\Delta\overrightarrow{E}_{(m)}-\Delta\overrightarrow{E}_{(\infty)} & =\overleftrightarrow{A}^{m}\cdot(\Delta\overrightarrow{E}_{(0)}-\Delta\overrightarrow{E}_{(\infty)})\nonumber \\
 & =\overleftrightarrow{A}^{m}\cdot(\Delta\overrightarrow{E}^{\prime}-\Delta\overrightarrow{E}_{(\infty)})\label{eq:derive_convergence_2}
\end{align}
where we used $\Delta\overrightarrow{E}_{(0)}=\Delta\overrightarrow{E}^{\prime}$
in the last step. Multiply Eq(\ref{eq:derive_convergence_2}) on the
left by $(\mathbb{1}-\overleftrightarrow{A})$ and use Eq(\ref{eq:derive_convergence_1}),
we obtain
\begin{equation}
(\mathbb{1}-\overleftrightarrow{A})\cdot\Delta\overrightarrow{E}_{(m)}-\Delta\overrightarrow{E}^{\prime}=(\mathbb{1}-\overleftrightarrow{A})\cdot\overleftrightarrow{A}^{m}\Delta\overrightarrow{E}^{\prime}-\overleftrightarrow{A}^{m}\Delta\overrightarrow{E}^{\prime}
\end{equation}
Thus
\begin{equation}
\Delta\overrightarrow{E}_{(m)}=(\mathbb{1}-\overleftrightarrow{A})^{-1}\cdot(\mathbb{1}-\overleftrightarrow{A}^{m+1})\cdot\Delta\overrightarrow{E}^{\prime}\label{eq:derive_convergence_3}
\end{equation}
Because the eigenvalue of matrix $\overleftrightarrow{A}$ has $|\lambda|<1$,
Eq(\ref{eq:derive_convergence_3}) always converge to $(\mathbb{1}-\overleftrightarrow{A})^{-1}\cdot\Delta\overrightarrow{E}^{\prime}$
with $m\to\infty$. Since Eq(\ref{eq:derive_convergence_3}) holds
for each time step, we have 
\begin{equation}
\overrightarrow{E}_{(m)}=(\mathbb{1}-\overleftrightarrow{A})^{-1}\cdot(\mathbb{1}-\overleftrightarrow{A}^{m+1})\cdot\overrightarrow{E}^{\prime}\label{eq:derive_convergence_4}
\end{equation}
From the Fourier transform of Eqs(\ref{eq:ampere_aux_x}) to (\ref{eq:ampere_aux_z}),
we obtain
\begin{equation}
\overrightarrow{E}^{\prime}=-c\frac{[\overrightarrow{k}]\times(\mathscr{A}_{BS}\overrightarrow{B})}{[\omega_{(m)}]}\label{eq:derive_convergence_5}
\end{equation}
where 
\begin{equation}
\mathscr{A}_{BS}=\begin{pmatrix}\alpha_{x} & 0 & 0\\
0 & \alpha_{y} & 0\\
0 & 0 & \alpha_{z}
\end{pmatrix}
\end{equation}
is the spatial averaging operator and $\omega_{(m)}$ is the frequency
of electromagnetic wave with $m$ iterations. In Eqs(\ref{eq:faraday_x})
to \ref{eq:faraday_z}, we use $\overrightarrow{E}_{(m)}$ as an approximation
for $\overrightarrow{E}_{(\infty)}$ to update $\overrightarrow{B}$,
thus

\begin{equation}
[\overrightarrow{k}]\times\overrightarrow{E}_{(m)}=\frac{1}{c}[\omega_{(m)}]\overrightarrow{B}\label{eq:derive_convergence_6}
\end{equation}
From Eqs(\ref{eq:derive_convergence_4}) and (\ref{eq:derive_convergence_5})
and (\ref{eq:derive_convergence_6}) we can eliminate $\overrightarrow{B}$
and $\overrightarrow{E}^{\prime}$ and get the equation for $\overrightarrow{E}_{(m)}$
\begin{equation}
[\omega_{(m)}]^{2}(\mathbb{1}-\overleftrightarrow{A})^{-1}\cdot(\mathbb{1}-\overleftrightarrow{A}^{m+1})\cdot\overrightarrow{E}_{(m)}+c^{2}[\overrightarrow{k}]\times(\mathscr{A}_{BS}[\overrightarrow{k}]\times\overrightarrow{E}_{(m)})\label{eq:derive_convergence_7}
\end{equation}
For $m\to\infty$
\begin{equation}
[\omega_{(\infty)}]^{2}(\mathbb{1}-\overleftrightarrow{A})^{-1}\cdot\overrightarrow{E}_{(\infty)}+c^{2}[\overrightarrow{k}]\times(\mathscr{A}_{BS}[\overrightarrow{k}]\times\overrightarrow{E}_{(\infty)})\label{eq:derive_convergence_8}
\end{equation}
If $\overrightarrow{E}_{(m)}$ is a eigenvector of $\overleftrightarrow{A}$
with eigenvalue $\lambda$, then from Eq(\ref{eq:derive_convergence_4})
we know that both $\overrightarrow{E}_{(\infty)}$ and $\overrightarrow{E}_{(0)}$
are eigenvectors of $\overleftrightarrow{A}$ with eigenvalue $\lambda$.
By comparing Eq(\ref{eq:derive_convergence_7}) and Eq(\ref{eq:derive_convergence_8})
we obtain
\begin{equation}
[\omega_{(m)}]^{2}(1-\lambda^{m+1})=[\omega_{(\infty)}]^{2}\label{eq:derive_convergence_9}
\end{equation}
where $\omega_{(\infty)}$ is the frequency of electromagnetic wave
with infinite number of iterations, which is given by Eq(\ref{eq:dispersion-em-original}).
We can rewrite Eq(\ref{eq:derive_convergence_9}) as

\begin{equation}
\bigg|\frac{[\omega_{(\infty)}]^{2}}{[\omega_{(m)}]^{2}}-1\bigg|=\bigg|\lambda\bigg|^{m+1}
\end{equation}
With sufficient number of iterations, $\omega_{(m)}$ is close to
$\omega_{(\infty)}$, thus
\begin{align}
\bigg|\frac{[\omega_{(m)}]^{2}}{[\omega_{(\infty)}]^{2}}-1\bigg| & =\bigg|\frac{[\omega_{(m)}]^{2}-[\omega_{(\infty)}]^{2}}{[\omega_{(\infty)}]^{2}}\bigg|\nonumber \\
 & \approx\bigg|\frac{[\omega_{(\infty)}]^{2}-[\omega_{(m)}]^{2}}{[\omega_{(m)}]^{2}}\bigg|\nonumber \\
 & =\bigg|\lambda\bigg|^{m+1}
\end{align}
For each set of $(k_{x},k_{y},k_{z})$, $\overleftrightarrow{A}$
has three eigenvalues $\lambda_{1},\lambda_{2},\lambda_{3}$. And
$\lambda_{1},\lambda_{2},\lambda_{3}$ are functions of $(k_{x},k_{y},k_{z})$.
Let $\lambda_{3}$ be the eigenvalue with the largest absolute value
among $\lambda_{1},\lambda_{2},\lambda_{3}$, i.e. $|\lambda_{1}|\le|\lambda_{2}|\le|\lambda_{3}|$,
then with sufficient number of iterations the dominant error term
for dispersion relation is 
\begin{equation}
\bigg|\frac{[\omega_{(m)}]^{2}}{[\omega_{(\infty)}]^{2}}-1\bigg|\approx\bigg|\lambda_{3}\bigg|^{m+1}\label{eq:dispersion_error}
\end{equation}

\section{Asymptotic NCI growth rate\label{sec:Derivation-of-NCI}}

Following the derivation in Xu 2013\citep{NCI_Xu2013}, we derive
the asymptotic expression for NCI growth rate in 2D. We use the relativistic
electron plasma frequency $\omega_{p}=\sqrt{\frac{4\pi q^{2}n_{e}}{\gamma m_{e}}}$
instead of the expression $\omega_{p}=\sqrt{\frac{4\pi q^{2}n_{e}}{m_{e}}}$.
The definition of $\omega_{p}$ with $\gamma$ has physical meaning
of the oscillation frequency in cold drifting plasma. Eq(10) in Xu
2013 and Eq(\ref{eq:EJ}) in this paper are the same for the R.H.S,
but on the L.H.S of Eq(\ref{eq:EJ}) in this paper, the vacuum dielectric
tensor are different from Xu 2013 and is written down in Eqs(\ref{eq:eps_em_xx})
to (\ref{eq:eps_em_zx}). In this paper we have $[\overrightarrow{k}]_{E}=[\overrightarrow{k}]_{B}=[\overrightarrow{k}]$
written down in Eq(\ref{eq:braket_k}). The dielectric tensor for
the current part of drifting plasma is same as that in Xu 2013. We
ignore high order terms in $1/\gamma$, in $v_{0}\to c$ limit
\begin{align}
\epsilon_{xx}^{(J)} & =\omega_{p}^{2}\sum_{\mu,\vec{\nu}}(-1)^{\mu}\frac{S_{J_{x}}S_{B_{z}}[k]_{y}k_{y}^{\prime}c^{2}}{(\omega^{\prime}-ck_{x}^{\prime})^{2}}\\
\epsilon_{xy}^{(J)} & =\omega_{p}^{2}\sum_{\mu,\vec{\nu}}(-1)^{\mu}\frac{k_{y}^{\prime}S_{J_{x}}(S_{E_{y}}[\omega]-S_{B_{z}}[k]_{x}c)c}{(\omega^{\prime}-ck_{x}^{\prime})^{2}}\\
\epsilon_{yx}^{(J)} & =\omega_{p}^{2}\sum_{\mu,\vec{\nu}}(-1)^{\mu}\frac{S_{J_{y}}S_{B_{z}}[k]_{y}c}{\omega^{\prime}-ck_{x}^{\prime}}\\
\epsilon_{yy}^{(J)} & =\omega_{p}^{2}\sum_{\mu,\vec{\nu}}(-1)^{\mu}\frac{S_{J_{y}}(S_{E_{y}}[\omega]-S_{B_{z}}[k]_{x}c)}{\omega^{\prime}-ck_{x}^{\prime}}\\
\epsilon_{zz}^{(J)} & =\omega_{p}^{2}\sum_{\mu,\vec{\nu}}(-1)^{\mu}\frac{S_{J_{z}}(S_{E_{z}}[\omega]-S_{B_{y}}[k]_{x}c)}{\omega^{\prime}-ck_{x}^{\prime}}
\end{align}
where
\begin{align}
\omega^{\prime}=\omega+\mu\frac{2\pi}{\Delta t} & \qquad\mu=0,\pm1,\pm2,\dots\\
k_{x}^{\prime}=k_{x}+\nu_{x}\frac{2\pi}{\Delta x} & \qquad\nu_{x}=0,\pm1,\pm2,\dots\label{eq:aliasing_x}\\
k_{y}^{\prime}=k_{y}+\nu_{y}\frac{2\pi}{\Delta y} & \qquad\nu_{y}=0,\pm1,\pm2,\dots
\end{align}
are the aliased frequency and wavevector, the interpolation functions
have dummy variables with aliasing wavevector. We expand $\omega^{\prime}$
around the beam resonance $\omega^{\prime}=k_{x}^{\prime}$ and write
$\omega^{\prime}=k_{x}^{\prime}+\delta\omega^{\prime}$, where $\delta\omega^{\prime}$
is a small term. Denoting $Q_{ij}$ as the numerator $\epsilon_{ij}^{(J)}$
we can write $\epsilon_{ij}=\epsilon_{ij}^{(\mathrm{vac})}+\epsilon_{ij}^{(J)}$
as
\begin{align}
\epsilon_{xx} & =\alpha_{y}[\omega]^{2}-\alpha_{t}c^{2}[k]_{y}^{2}-\frac{Q_{xx}}{\delta\omega^{\prime2}}\\
\epsilon_{xy} & =\alpha_{t}c^{2}[k]_{x}[k]_{y}-\frac{Q_{xy}}{\delta\omega^{\prime2}}\\
\epsilon_{yx} & =\alpha_{t}c^{2}[k]_{x}[k]_{y}-\frac{Q_{yx}}{\delta\omega^{\prime}}\\
\epsilon_{yy} & =\alpha_{x}[\omega]^{2}-\alpha_{t}c^{2}[k_{x}]^{2}-\frac{Q_{yy}}{\delta\omega^{\prime}}
\end{align}
Using $\det(\epsilon)=0$ and only keeping $x$ and $y$ terms, we
can obtain
\begin{equation}
A_{1}\delta\omega^{\prime2}+B_{1}\delta\omega^{\prime}+C_{1}=0
\end{equation}
where 
\begin{align}
A_{1} & =\alpha_{x}\alpha_{y}[\omega]^{4}-\alpha_{t}[\omega]^{2}c^{2}(\alpha_{y}[k]_{x}^{2}+\alpha_{x}[k]_{y}^{2})\nonumber \\
 & =[\omega]^{2}(\alpha_{x}\alpha_{y}[\omega]^{2}-\alpha_{t}\alpha_{y}c^{2}[k]_{x}^{2}-\alpha_{t}\alpha_{x}c^{2}[k]_{y}^{2})\\
B_{1} & =\alpha_{t}c^{2}[k]_{x}[k]_{y}Q_{yx}-(\alpha_{y}[\omega]^{2}-\alpha_{t}c^{2}[k]_{y}^{2})Q_{yy}\\
C_{1} & =-(\alpha_{x}[\omega]^{2}-\alpha_{t}c^{2}[k]_{x}^{2})Q_{xx}+\alpha_{t}c^{2}[k]_{x}[k]_{y}Q_{xy}
\end{align}
Now we use the condition that $(\omega^{\prime},k_{x}^{\prime})$
sits near the EM modes
\begin{equation}
\alpha_{t}^{-1}\frac{[\omega]^{2}}{c^{2}}=\alpha_{x}^{-1}[k_{x}]^{2}+\alpha_{y}^{-1}[k_{y}]^{2}
\end{equation}
and expand the finite difference operator $[\omega]=\xi_{0}+\delta\omega^{\prime}\xi_{1}$
, and $\alpha_{t}=\xi_{2}+\delta\omega^{\prime}\xi_{3}$ where 
\begin{align}
\xi_{0} & =\frac{\sin(\tilde{k}c\Delta t/2)}{\Delta t/2}\label{eq:xi0}\\
\xi_{1} & =\cos(\tilde{k}\Delta t/2)\\
\xi_{2} & =\frac{2+\cos(\tilde{k}c\Delta t)}{3}\\
\xi_{3} & =-\frac{\Delta t\sin(\tilde{k}c\Delta t)}{3}\label{eq:xi3}
\end{align}
\begin{equation}
\tilde{k}=k_{x}+\nu_{x}\frac{2\pi}{\Delta x}-\mu\frac{2\pi}{c\Delta t}
\end{equation}
and then we further expand $[\omega]$ to first order in $A_{1}$,
since this term is sensitive near the EM mode, while only keeping
the zeroth order of $[\omega]$ in $B_{1}$ and $C_{1}$
\begin{align}
A_{1} & =\xi_{0}^{2}(2\alpha_{x}\alpha_{y}\xi_{0}\xi_{1}-\xi_{3}\alpha_{y}[k]_{x}^{2}-\xi_{3}\alpha_{x}[k]_{y}^{2})\delta\omega^{\prime}\nonumber \\
 & =\xi_{0}^{2}(2\alpha_{x}\alpha_{y}\xi_{0}\xi_{1}-\xi_{3}\alpha_{x}\alpha_{y}\alpha_{t}^{-1}[\omega]^{2})\delta\omega^{\prime}\nonumber \\
 & \approx\xi_{0}^{2}(2\alpha_{x}\alpha_{y}\xi_{0}\xi_{1}-\xi_{3}\alpha_{x}\alpha_{y}\xi_{2}^{-1}\xi_{0}^{2})\delta\omega^{\prime}\\
B_{1} & =\xi_{2}c^{2}[k]_{x}[k]_{y}Q_{yx}-(\alpha_{y}\xi_{0}^{2}-\xi_{2}c^{2}[k]_{y}^{2})Q_{yy}\\
C_{1} & =-(\alpha_{x}\xi_{0}^{2}-\xi_{2}c^{2}[k]_{x}^{2})Q_{xx}+\xi_{2}c^{2}[k]_{x}[k]_{y}Q_{xy}
\end{align}
We then obtain a cubic equation, we can drop the small $B_{1}$ term
\begin{align}
 & \xi_{0}^{2}(2\alpha_{x}\alpha_{y}\xi_{0}\xi_{1}-\xi_{3}\alpha_{x}\alpha_{y}\xi_{2}^{-1}\xi_{0}^{2})\delta\omega^{\prime3}\nonumber \\
 & \qquad-(\alpha_{x}\xi_{0}^{2}-\xi_{2}c^{2}[k]_{x}^{2})Q_{xx}+\xi_{2}c^{2}[k]_{x}[k]_{y}Q_{xy}
\end{align}
So the growth rate
\begin{align}
\Gamma(\vec{k}) & =\frac{\sqrt{3}}{2}\bigg|\frac{(\alpha_{x}\xi_{0}^{2}-\xi_{2}c^{2}[k]_{x}^{2})Q_{xx}-\xi_{2}c^{2}[k]_{x}[k]_{y}Q_{xy}}{\xi_{0}^{2}(2\alpha_{x}\alpha_{y}\xi_{0}\xi_{1}-\xi_{3}\alpha_{x}\alpha_{y}\xi_{2}^{-1}\xi_{0}^{2})}\bigg|^{1/3}\nonumber \\
 & =\frac{\sqrt{3}}{2}\bigg|\frac{(\alpha_{x}\xi_{0}^{2}-\xi_{2}c^{2}[k]_{x}^{2})Q_{xx}-\xi_{2}c^{2}[k]_{x}[k]_{y}Q_{xy}}{\xi_{0}^{3}\alpha_{x}\alpha_{y}(2\xi_{1}-\xi_{3}\xi_{2}^{-1}\xi_{0})}\bigg|^{1/3}\nonumber \\
 & =\frac{\sqrt{3}}{2}\bigg|\frac{\omega_{p}^{2}c^{2}[k]_{y}k_{y}S_{J_{x}}(\alpha_{x}\xi_{0}S_{B_{z}}-\xi_{2}c[k]_{x}S_{E_{y}})}{\xi_{0}^{2}\alpha_{x}\alpha_{y}(2\xi_{1}-\xi_{3}\xi_{2}^{-1}\xi_{0})}\bigg|^{1/3}
\end{align}

\clearpage{}

\bibliographystyle{plainnat}
\bibliography{pic_method}

\end{document}